\documentclass[iop]{emulateapj}

\pdfoutput=1

\usepackage{amsmath}
\usepackage{amssymb}
\usepackage{apjfonts}
\usepackage{color}
\usepackage{graphicx}
\usepackage{hyperref}
\usepackage{multirow}

\hypersetup{colorlinks,
  linkcolor=blue,
  filecolor=blue,
  urlcolor=blue,
  citecolor=blue}

\def\chandra{{\it Chandra}}
\def\planck{{\it Planck}}
\def\spitzer{{\it Spitzer}}

\def\rosat{{\it ROSAT}}
\def\xmm{{\it XMM-Newton}}
\def\wmap{{\it WMAP}}

\def\jcap{JCAP}
\def\msngr{Msngr}
\def\phrep{PhR}
\def\prd{PhRvD}
\def\rmp{RvMP}

\slugcomment{Accepted for publication in The Astrophysical Journal}
\submitted{Accepted for publication in The Astrophysical Journal}

\shorttitle{ACT Dynamical Masses of Clusters}
\shortauthors{C.~Sif\'on et al.}

\def\ytslope{$0.79\pm0.15$}
\def\ycslope{$0.84\pm0.14$}
\def\Yslope{$0.48\pm0.11$}
\def\ytzero{$15.02\pm0.07$}
\def\yczero{$15.02\pm0.07$}
\def\Yzero{$14.99\pm0.07$}
\def\ytscatter{$0.13\pm0.05$}
\def\ycscatter{$0.13\pm0.05$}
\def\Yscatter{$0.18\pm0.10$}

\begin{document}

\title{The Atacama Cosmology Telescope: Dynamical Masses and Scaling Relations
       for a Sample of Massive Sunyaev-Zel'dovich Effect Selected Galaxy
       Clusters\footnotemark[\dag]\footnotemark[\ddag]}

\author{
Crist\'obal~Sif\'on\altaffilmark{1,2,21}, % PUC, Leiden
Felipe~Menanteau\altaffilmark{3,21}, % Rutgers
Matthew~Hasselfield\altaffilmark{4}, % UBC
Tobias~A.~Marriage\altaffilmark{5}, % Hopkins
John~P.~Hughes\altaffilmark{3,21}, % Rutgers
L.~Felipe~Barrientos\altaffilmark{1}, % PUC
Jorge~Gonz\'alez\altaffilmark{1}, % PUC
Leopoldo~Infante\altaffilmark{1}, % PUC
Graeme~E.~Addison\altaffilmark{6}, % Oxford
Andrew~J.~Baker\altaffilmark{3}, % Rutgers
Nick~Battaglia\altaffilmark{7}, % CITA
J.~Richard~Bond\altaffilmark{7}, % CITA
Devin~Crichton\altaffilmark{5}, % Hopkins
Sudeep~Das\altaffilmark{8}, % Berkeley
Mark~J.~Devlin\altaffilmark{9}, % UPenn
Joanna~Dunkley\altaffilmark{6}, % Oxford
Rolando~D\"unner\altaffilmark{1}, % PUC
Megan~B.~Gralla\altaffilmark{5}, % Hopkins
Amir~Hajian\altaffilmark{7}, % CITA
Matt~Hilton\altaffilmark{10}, % Nottingham
Adam~D.~Hincks\altaffilmark{7,15}, % CITA, Jadwin
Arthur~B.~Kosowsky\altaffilmark{11}, % Pittsburgh
Danica~Marsden\altaffilmark{12}, % UCSB
Kavilan~Moodley\altaffilmark{13}, % KwaZulu-Natal
Michael~D.~Niemack\altaffilmark{14}, % Boulder
Michael~R.~Nolta\altaffilmark{7}, % CITA
Lyman~A.~Page\altaffilmark{15}, % Jadwin
Bruce~Partridge\altaffilmark{16}, % Haverford
Erik~D.~Reese\altaffilmark{9}, % UPenn
Neelima~Sehgal\altaffilmark{17}, % Peyton
Jon~Sievers\altaffilmark{7}, % CITA
David~N.~Spergel\altaffilmark{17}, % Peyton
Suzanne~T.~Staggs\altaffilmark{15}, % Jadwin
Robert~J.~Thornton\altaffilmark{18,9}, % West Chester, UPenn
Hy~Trac\altaffilmark{19}, % Carnegie-Pittsburgh
Edward~J.~Wollack\altaffilmark{20} % NASA/Goddard
}

\affil{$^{1}$Departamento de Astronom\'ia y Astrof\'isica, Facultad de F\'isica,
             Pontificia Universidad Cat\'olica de Chile, Casilla 306, Santiago
             22, Chile}
\affil{$^{2}$Leiden Observatory, Leiden University, PO Box 9513,
             NL-2300 RA Leiden, Netherlands}
\affil{$^{3}$Department of Physics and Astronomy, Rutgers University, 136 Frelinghuysen Rd,
             Piscataway, NJ 08854, USA}
\affil{$^{4}$Department of Physics and Astronomy, University of British Columbia,
             Vancouver, BC V6T 1Z4, Canada}
\affil{$^{5}$Department of Physics and Astronomy, The Johns Hopkins University,
             Baltimore, MD 21218-2686, USA}
\affil{$^{6}$Sub-department of Astrophysics, University of Oxford, Denys Wilkinson Building,
             Keble Road, Oxford OX1 3RH, UK}
\affil{$^{7}$Canadian Institute for Theoretical Astrophysics, University of Toronto,
             Toronto, ON M5S 3H8, Canada}
\affil{$^{8}$Berkeley Center for Cosmological Physics, LBL and Department of Physics,
             University of California, Berkeley, CA 94720, USA}
\affil{$^{9}$Department of Physics and Astronomy, University of Pennsylvania,
             209 South 33rd Street, Philadelphia, PA 19104, USA}
\affil{$^{10}$School of Physics and Astronomy, University of Nottingham,
              University Park, Nottingham, NG7 2RD, UK}
\affil{$^{11}$Physics and Astronomy Department, University of Pittsburgh,
              100 Allen Hall, 3941 O'Hara Street, Pittsburgh, PA 15260, USA}
\affil{$^{12}$Department of Physics, University of California-Santa Barbara,
              Santa Barbara, CA 93106-9530, USA}
\affil{$^{13}$University of KwaZulu-Natal, Astrophysics and Cosmology Research
              Unit, School of Mathematical Sciences, Durban, 4041, South Africa}
\affil{$^{14}$NIST Quantum Devices Group, 325 Broadway Mailcode 817.03,
              Boulder CO 80305, USA}
\affil{$^{15}$Joseph Henry Laboratories of Physics, Jadwin Hall, Princeton University,
              Princeton, NJ, 08544, USA}
\affil{$^{16}$Department of Physics and Astronomy, Haverford College, Haverford,
              PA 19041, USA}
\affil{$^{17}$Department of Astrophysical Sciences, Peyton Hall, Princeton University,
              Princeton, NJ, 08544, USA}
\affil{$^{18}$Department of Physics, West Chester University, West Chester,
              PA 19383, USA}
\affil{$^{19}$Department of Physics, Carnegie Mellon University, Pittsburgh,
              PA 15213, USA}
\affil{$^{20}$NASA/Goddard Space Flight Center,
              Greenbelt, MD 20771, USA}

\footnotetext[\dag]{Based in part on observations collected at the European Organisation for
Astronomical Research in the Southern Hemisphere, Chile, under programs 084.A-0577 and 086.A-0425.}
\footnotetext[\ddag]{Based in part on observations obtained at the Gemini Observatory, which is
operated by the Association of Universities for Research in Astronomy, Inc., under a cooperative
agreement with the NSF on behalf of the Gemini partnership: the National Science Foundation (United
States), the Science and Technology Facilities Council (United Kingdom), the National Research
Council (Canada), CONICYT (Chile), the Australian Research Council (Australia), Minist\'erio da
Ci\^encia e Tecnologia (Brazil) and Ministerio de Ciencia, Tecnolog\'ia e Innovaci\'on Productiva
(Argentina).}
\footnotetext[21]{Visiting Astronomer, Gemini South Observatory.}

\begin{abstract}

We present the first dynamical mass estimates and scaling relations for a sample of
Sunyaev--Zel'dovich Effect (SZE) selected galaxy clusters. The sample consists of 16 massive
clusters detected with the Atacama Cosmology Telescope (ACT) over a 455 $\mathrm{deg^2}$ area of the
southern sky.  Deep multi-object spectroscopic observations were taken to secure
intermediate-resolution ($R\sim700$--$800$) spectra and redshifts for $\approx60$ member galaxies on
average per cluster. The dynamical masses $M_{200c}$ of the clusters have been calculated using
simulation-based scaling relations between velocity dispersion and mass. The sample has a median
redshift $z = 0.50$ and a median mass $M_{200c}\simeq12\times10^{14}\,h_{70}^{-1}\,M_\odot$ with a
lower limit $M_{200c}\simeq6\times10^{14}\,h_{70}^{-1}\,M_\odot$, consistent with the expectations
for the ACT southern sky survey. These masses are compared to the ACT SZE properties of the sample,
specifically, the match-filtered central SZE amplitude $\widetilde{y_0}$, the central Compton
parameter $y_0$, and the integrated Compton signal $Y_{200c}$, which we use to derive SZE-mass
scaling relations. All SZE estimators correlate with dynamical mass with low intrinsic scatter
($\lesssim20\%$), in agreement with numerical simulations. We explore the effects of various
systematic effects on these scaling relations, including the correlation between observables and the
influence of dynamically disturbed clusters. Using the three-dimensional information available, we
divide the sample into relaxed and disturbed clusters and find that $\sim50\%$ of the clusters are
disturbed. There are hints that disturbed systems might bias the scaling relations but given the
current sample sizes these differences are not significant; further studies including more clusters
are required to assess the impact of these clusters on the scaling relations.

\end{abstract}


\section{Introduction}\label{sec:intro}

Studies of clusters of galaxies have had a wide impact on our understanding of galaxy formation and
cosmology \citep[see][for a review]{Voit-05}. They are a unique laboratory for studying the effects
of the environment (high density, gas pressure, collisions, etc.) on galaxy evolution
\citep{Butcher-84, Balogh-99, Hansen-09}. At the same time, number counts of galaxy clusters,
sensitive to the amplitude of matter fluctuations, can provide constraints on various cosmological
parameters \citep{Bahcall-98, Evrard-02, Vikhlinin-09, Mantz-10b, Mantz-10c, Rozo-10}. An accurate
determination of the latter requires that we know the mass and redshift distributions of clusters
with good precision.

The Sunyaev--Zel'dovich effect \citep[SZE;][]{Zeldovich-69,Sunyaev-70} is a distortion in the cosmic
microwave background (CMB) temperature produced by inverse-Compton scattering of CMB photons as they
interact with the hot electrons of the intracluster medium (ICM) of a galaxy cluster. Its surface
brightness is independent of redshift, and its strength is proportional to the line-of-sight
(l.o.s.) column density times the electron temperature. The SZE is a powerful tool for detecting
massive clusters to high redshifts \citep[see, e.g., the reviews by][]{Birkinshaw-99, Carlstrom-02}.

Early measurements of the SZE were achieved with targeted observations of known clusters. These
revealed the power of SZE studies, reaching from gas physics and inner structure of clusters
\citep{Grego-01, Benson-04}, to cosmological parameters such as the Hubble constant
\citep{Birkinshaw-91, Hughes-98} and the energy density of matter in the universe, $\Omega_M$
\citep{Grego-01}. Large SZE surveys over cosmologically significant areas of the sky have recently
come to fruition as the Atacama Cosmology Telescope \citep[ACT,][]{Fowler-07, Swetz-ACT} and the
South Pole Telescope \citep[SPT,][]{Carlstrom-SPT} have begun scanning large areas of the sky at
millimeter wavelengths. The \planck\ satellite \citep{Planck-pre} is conducting an all-sky survey
and has recently released the first all-sky sample of SZE-selected galaxy clusters
\citep{Planck-clusters}. The first cluster detections with ACT and SPT are presented in
\cite{Hincks-ACT} and \cite{Staniszewski-SPT}, respectively.

The rapidly growing SZE cluster samples have the potential to place strong constraints on
cosmological parameters \citep[e.g.,][]{Battye-03}. Both numerical simulations \citep{Springel-01a,
daSilva-04, Motl-05, Nagai-06, Battaglia-12} and analytical studies \citep{Reid-06, Ashfordi-08,
Shaw-08} suggest a tight correlation between cluster mass and SZE signal. On the other hand, biased
mass estimates can have a large impact on cosmological parameter determination
\citep[e.g.,][]{Francis-05}. By limiting their study to the high-significance clusters,
\cite{Sehgal-ACT} have shown the power of the ACT sample in constraining cosmological parameters,
particularly the dark energy equation-of-state parameter $w$ and the root-mean-square (rms) mass
fluctuations on a scale of $8\,h^{-1}\,\rm{Mpc}$, $\sigma_8$. Likewise, \cite{Vanderlinde-SPT} have
used SPT data to set cosmological constraints, with similar findings. They have also shown that
these improvements can be achieved only in the presence of a well-calibrated scaling relation
between mass and SZE signal. To assess the scaling of SZE signal with mass, independent means of
measuring the mass are crucial.

\cite{Benson-13} used X-ray observations in combination with SZE measurements to derive an empirical
scaling relation between mass and SZE signal. This allowed them to confirm that SZE-selected samples
of clusters yield significant improvements when added to other datasets to constrain cosmological
parameters. While X-ray observations have proven to be an effective way of measuring cluster masses,
and have been exploited to characterize the SZE signal \citep{Laroque-06, Bonamente-08,
Andersson-SPT, Melin-11, Planck-scaling}, they do not provide truly independent mass estimates from
SZE measurements, since both rely on the properties of the gas in the ICM and should be affected by
similar physics.

The velocity dispersion of cluster member galaxies is one of the most widely used methods for
constraining cluster mass, and is independent of the properties of the gas in the ICM. It takes into
account the galaxy distribution and relies, to some extent, on the assumption that the clusters are
relaxed (i.e., virialized). Until recently, however, mass measurements to independently calibrate
the SZE signal with mass have come from optical richness \citep{Menanteau-09, High-SPT,  
Menanteau-10, Planck-optical} and lensing analyses \citep{Sealfon-06, Umetsu-09, Marrone-12}.
\cite{Hand-ACT} presented stacked ACT data in the directions of luminous red galaxies from
the Sloan Digital Sky Survey (SDSS) Data Release 7 \citep[DR7,][]{SDSS-DR7} using optical
luminosity-based masses. This approach allowed them to probe the SZE signal from lower mass systems
than otherwise possible.

\cite{Rines-10} presented the first statistical comparison between dynamically estimated masses and
integrated SZE signal from a sample of 15 nearby ($z < 0.3$) galaxy clusters, showing that masses
thus determined and the integrated SZE flux are correlated at the $\approx99\%$ confidence level.
Furthermore, they estimate that the significance is higher than that of the correlation between SZE
and weak lensing masses from \cite{Marrone-09}, probably because of the smaller apertures used in
the
latter study. However, since their sample was not homogeneously selected, \cite{Rines-10} do not
account for observational biases in their sample and do not report a formal scaling relation between
mass and SZE flux.

In this work we present spectroscopic redshifts and---for the first time---dynamical masses of a
sample of clusters of galaxies selected with the SZE. These clusters were observed by ACT in its
2008 southern sky survey at 148 GHz \citep{Marriage-ACT}, and were optically confirmed by
\cite{Menanteau-ACT}. We use a variety of SZE diagnostics to assess the scaling with dynamical mass
and thus present the first robust scaling relations between dynamical masses and SZE signal for a
sample of SZE selected clusters.

Throughout this work we use a flat $\Lambda$CDM cosmology consistent with \wmap-7 data
\citep{Komatsu-WMAP7}, with $\Omega_\Lambda = 0.73$, $\Omega_M = 0.27$ and $H_0 =
70h_{70}\,\mathrm{km\,s^{-1}\,Mpc^{-1}}$. Masses and integrated SZE signals are estimated within a
radius $r_{200c}$ which encloses a density 200 times the critical density of the universe at the
redshift of the cluster, $\rho_c(z) = 3H^2(z)/8\pi G$. All quoted errors are 68\% confidence
intervals unless otherwise stated.

\section{Observations}

\subsection{ACT SZE Observations}\label{sec:sze}

ACT is a 6 m off-axis Gregorian telescope operating at an altitude of 5200 m in the Atacama
Desert in Chile, designed to observe the CMB at arcminute-scale resolution. It has three
1024-element arrays of transition edge sensors operating at 148, 218 and 277 GHz. ACT surveyed two
regions of the sky, of which 755 $\mathrm{deg^2}$ have been used for cluster studies
\citep{Marriage-ACTb, Hasselfield-ACT}. The processes of cluster detection and extraction are
thoroughly described in \cite{Marriage-ACT}, and references therein. In short, the maps are
match-filtered and convolved with a beta-model profile with $\beta = 0.86$ with varying core radius
$\theta_c$ from $0.\!\arcmin25$ to $4.\!\arcmin0$. Cluster signal-to-noise ratio (S/N) is
measured as the maximum S/N from this set of filtered maps.

We report on a large spectroscopic follow-up campaign of an ACT 148 GHz cluster sample, which was
obtained from a 455 $\mathrm{deg^2}$ survey of the southern sky. The survey is roughly bounded by
right ascensions $00^{\rm h}12^{\rm m}$ and $07^{\rm h}08^{\rm m}$ and declinations
$-56^\circ11\arcmin$ and $-49^\circ00\arcmin$. For further details on the ACT observations, map
making, data reduction, and cluster detection procedure, see
\cite{Fowler-10}, \cite{Marriage-ACT}, and \cite{Dunner-ACT}.

\subsubsection{The Cluster Sample}\label{sec:sample}

In this study we consider a total of 19 clusters, spanning a wide range in mass and redshift. We
focus, in particular, on the subsample of 16 clusters that were detected by ACT through their SZ
signal. This subsample contains 15 systems that were detected by ACT in the 2008 single-season maps
\citep{Marriage-ACT} and confirmed optically on 4m-class telescopes \citep{Menanteau-ACT}, plus one
additional cluster (ACT-CL~J0521$-$5104) detected in the new analysis of multi-season maps. This
latter cluster was initially targeted for spectroscopic follow-up based on its optical richness
alone \citep{Menanteau-10}. The 16 clusters were selected based on a redshift cut of $z_{\rm  
phot}>0.35$ and were all discovered with the SZE, with the exceptions of ACT-CL~J0330$-$5227
\citep[X-rays,][]{Werner-07} and ACT-CL~J0521$-$5104 \citep[optical,][]{Menanteau-10}.
ACT-CL~J0330$-$5227 is located 12\arcmin\ northeast (NE) of A3128
\citep[$z=0.06$;][]{Colless-87,Katgert-96}, but \cite{Werner-07} found it to be an unrelated,
background cluster at $z=0.44$ based on the observed energy of the Fe K X-ray emission line using
\xmm\ observations and the optical spectrum of the brightest cluster galaxy (BCG). \cite{Hincks-ACT}
have shown that the observed SZE signal is clearly related to the background cluster. Four clusters
were initially reported by SPT \citep[][see Section
\ref{sec:previous}]{Staniszewski-SPT,Vanderlinde-SPT} and studied optically by \cite{Menanteau-10}. 
ACT-CL~J0546$-$5345 is the only cluster with a dynamical mass estimate prior to this study
\citep[][see Section \ref{sec:previous-J0546}]{Brodwin-SPT}. Recently, \cite{Hilton-13} presented a
study of the stellar content of 14 of these 16 clusters from \spitzer\ observations.

Thus of the 16 SZE-detected clusters reported here, 10 are newly discovered by ACT.
\cite{Menanteau-ACT} confirmed them as clusters with a BCG and an accompanying red sequence of
galaxies and studied their X-ray properties from archival \rosat\ data for the 15 clusters, plus
\chandra\ and/or \xmm\ data when available. The clusters cover the range $\sim1$---$30\times10^{44}
\,\mathrm{erg\,s^{-1}}$ in X-ray luminosity as measured in the 0.1--2.4 keV band. Photometric
redshifts were estimated for these 15 clusters by \cite{Menanteau-ACT}. The spectroscopic redshift
range covered by the sample is $0.28 < z < 1.07$ with a median redshift $z=0.50$.

Additionally, we report on three optically selected, high-richness galaxy clusters from the Southern
Cosmology Survey (SCS; \citealt{Menanteau-10}). These clusters were part of our 2009B follow-up
observations before the ACT maps were available for cluster detection, and were not detected by ACT.
They are briefly discussed in Section \ref{sec:SCS}.

\subsubsection{Cluster SZE Measurements}\label{sec:sze_par}

%%%%%%%%%%%%%%%%%%%%%%%%%%%
%%%   SZ effect Table   %%%
%%%%%%%%%%%%%%%%%%%%%%%%%%%
\def\tna{\tablenotemark{a}}
\def\tnb{\tablenotemark{b}}
\def\tnc{\tablenotemark{c}}
\begin{deluxetable}{ l c c c c }
\tablecaption{ACT-SZE Measurements of Clusters}
\tablehead{
\colhead{Cluster} &
\colhead{$z$} &
\colhead{$\widetilde{y_0}$\tna} &
\colhead{$y_0$\tnb} &
\colhead{$Y_{200c}$\tnc}
 \\
\colhead{} &
\colhead{} &
\colhead{$(10^{-4})$} &
\colhead{$(10^{-4})$} &
\colhead{$(10^{-10})$}
}
\startdata
ACT-CL~J0102$-$4915 & $0.870$ & $3.51\pm0.43$ & $5.66\pm0.62$ & $1.47\pm0.18$ \\
ACT-CL~J0215$-$5212 & $0.480$ & $0.78\pm0.18$ & $1.10\pm0.25$ & $0.37\pm0.10$ \\
ACT-CL~J0232$-$5257 & $0.556$ & $0.60\pm0.17$ & $0.91\pm0.28$ & $0.28\pm0.07$ \\
ACT-CL~J0235$-$5121 & $0.278$ & $0.99\pm0.19$ & $1.03\pm0.21$ & $0.97\pm0.20$ \\
ACT-CL~J0237$-$4939 & $0.334$ & $0.93\pm0.26$ & $0.94\pm0.32$ & $1.07\pm0.31$ \\
ACT-CL~J0304$-$4921 & $0.392$ & $1.59\pm0.31$ & $1.68\pm0.37$ & $1.05\pm0.25$ \\
ACT-CL~J0330$-$5227 & $0.442$ & $1.25\pm0.18$ & $1.61\pm0.21$ & $0.90\pm0.13$ \\
ACT-CL~J0346$-$5438 & $0.530$ & $1.05\pm0.22$ & $1.48\pm0.30$ & $0.46\pm0.11$ \\
ACT-CL~J0438$-$5419 & $0.421$ & $1.63\pm0.13$ & $2.06\pm0.15$ & $1.14\pm0.10$ \\
ACT-CL~J0509$-$5341 & $0.461$ & $0.82\pm0.14$ & $0.59\pm0.19$ & $0.12\pm0.05$ \\
ACT-CL~J0521$-$5104 & $0.675$ & $0.72\pm0.16$ & $1.31\pm0.25$ & $0.28\pm0.07$ \\
ACT-CL~J0528$-$5259 & $0.768$ & $0.49\pm0.13$ & $1.03\pm0.27$ & $0.10\pm0.03$ \\
ACT-CL~J0546$-$5345 & $1.066$ & $0.92\pm0.14$ & $2.36\pm0.30$ & $0.26\pm0.03$ \\
ACT-CL~J0559$-$5249 & $0.609$ & $0.90\pm0.14$ & $1.51\pm0.20$ & $0.51\pm0.05$ \\
ACT-CL~J0616$-$5227 & $0.684$ & $1.00\pm0.15$ & $1.80\pm0.22$ & $0.47\pm0.05$ \\
ACT-CL~J0707$-$5522 & $0.296$ & $0.52\pm0.21$ & $0.51\pm0.22$ & $0.57\pm0.13$
\enddata
\tablecomments{Redshifts are listed for reference. See Table \ref{t:dyn} for details.}
\tablenotetext{a}{Central match-filtered amplitude of the SZE, measured using the A10 profile with
                  a FWHM of 2\arcmin. See \cite{Hasselfield-ACT}.}
\tablenotetext{b}{Projected central Compton parameter assuming the A10 profile. See
                  \cite{Hasselfield-ACT}.}
\tablenotetext{c}{Spherically-integrated Compton amplitude within $r_{200c}$ assuming the A10
                  profile. See Section \ref{sec:sze_par}.}
\label{t:sze}
\end{deluxetable}

To characterize the SZE produced by each cluster (in the 148 GHz band) we study three different
estimators. These values are listed in Table \ref{t:sze} and are all measured using multi-season
(2008--2010) ACT data. 

The first estimator, $\widetilde{y_0}$, corresponds to the central match-filtered SZE amplitude.
A detailed description of the procedure is given in \citet[see their Section 2.2]{Hasselfield-ACT},
but is outlined here. The ACT maps are passed through a matched filter to extract the amplitude of
the temperature decrement of clusters modeled with the universal pressure profile of
\cite{Arnaud-10}---hereafter ``the A10 profile''---with a fixed scale $\theta_{\rm FWHM}=2\arcmin$.
This scale is related to the more usual parameterization of the characteristic scale by
$\theta_{500c} = 2.94\,\theta_{\rm FWHM}$ given the best-fit concentration parameter from
\cite{Arnaud-10}, $c_{500c}=1.177$. Although the filter accounts for the effects of the beam in the
signal template, its normalization is chosen to return the central decrement of the corresponding
unconvolved cluster signal. The central temperature decrement is scaled to a central Compton
parameter using the standard non-relativistic SZE frequency dependence \citep{Sunyaev-80}.

Using $\widetilde{y_0}$ and assuming that the pressure profile follows the (mass dependent) A10
profile, one can estimate what the actual central Compton parameter should be. While this measure
carries some assumptions about the physics of the cluster and the relation between pressure and mass
(i.e., it is model-dependent), it is completely independent from the reported dynamical masses and
it is thus still useful to compare both quantities. The central Compton parameter is referred to as
$y_0$, as usual. A more detailed discussion about $\widetilde{y_0}$ and $y_0$ can be found in
\cite{Hasselfield-ACT}.

Our third measurement is the integrated Compton signal. Large integration areas tend to give
measurements that are more robust against the effects of cluster physics such as active galactic
nucleus (AGN) feedback \citep{Motl-05,Nagai-06,Reid-06}, and to projection effects \citep{Shaw-08}.
Dynamical masses are usually measured at $r_{200c}$ (see Section \ref{sec:masses}), providing
therefore a measurement of the size of the cluster. Since the parameterization of the A10 profile is
given in terms of quantities measured at $r_{500c}$, we convert values from $r_{200c}$ to $r_{500c}$
using a \cite{NFW-95} profile (hereafter NFW profile) and the mass--concentration relation of
\cite{Duffy-08}. Combined with the dynamical information, this sets the scale of the filter through
$\theta_{500c}$. The filtering then returns the total integrated profile out to the virial radius,
which is scaled to $r_{200c}$ using the prescription of \cite{Arnaud-10}. We refer to these
spherical SZE measurements within $r_{200c}$ as $Y_{200c}$ hereafter. We estimate the covariance
between $Y_{200c}$ and $M_{200c}$ by measuring $Y(<r)$ from the maps at different radii around
$r_{200c}$ for each cluster; the dynamical mass is re-scaled assuming a spherical cluster. This
covariance is included in the determination of the scaling relations (see
Sections \ref{sec:fits} and \ref{sec:rcorr}).

\subsection{Optical Spectroscopy}

%%%%%%%%%%%%%%%%%%%%%%%%%%%%
%%%  Observations Table  %%%
%%%%%%%%%%%%%%%%%%%%%%%%%%%%
\begin{deluxetable*}{c c l l c c l c c}[b]
\tablecaption{Spectroscopic Observation Details}
\tablehead{
\colhead{Run} &
\colhead{Semester} &
\colhead{PI} &
\colhead{Tel./Inst.} &
\colhead{Program ID} &
\colhead{Mode} &
\colhead{Grating} &
\colhead{Hours} &
\colhead{$N_{\rm cl}$}
}
\startdata
1 & 2009B & Infante                  & VLT/FORS2     &  084.A-0577  & Service   & GRIS 300I+11 & 14
& 3  \\
2 & 2009B & Barrientos               & Gemini-S/GMOS & GS-2009B-Q-2 & Service   & R400\_G5325  & 20
& 4  \\
3 & 2010B & Infante                  & VLT/FORS2     &  086.A-0425  & Service   & GRIS 300I+11 & 15
& 2  \\
4 & 2010B & Barrientos/Menanteau\tna & Gemini-S/GMOS & GS-2010B-C-2 & Classical & R400\_G5325  & 40
& 10
\enddata
\tablecomments{$N_{\rm cl}$ is the number of clusters observed in each run. Each cluster was fully
               observed in one run.}
\tablenotetext{a}{Joint Chile/US proposal}
\label{t:runs}
\end{deluxetable*}

The spectroscopic observations were carried out in two semesters, 2009B and 2010B. Each semester was
split into two observing runs, one with FORS2 at the Very Large Telescope \citep[VLT;][]{FORS} and
one with GMOS at Gemini South \citep{GMOS}, both telescopes located in Chile. The details of each
observing run are listed in Table \ref{t:runs}. In total, we had 89 hr of observation, during
which we collected multi-object spectroscopy for 19 clusters.

Targets were selected by a two-step process. First, a photometric redshift-selected catalog was
constructed, including galaxies within $\pm0.1$ of the redshift of the BCG. Within this catalog,
galaxies were visually selected based on their $gri$ colors, with preference given to bright
galaxies. All our spectroscopic observations cover the wavelength range $\sim4000-8000$\AA. In this
range several spectral features are observable at the median photometric redshift of 0.54
\citep{Menanteau-ACT}. These are mainly the Ca {\sc ii} K--H absorption doublet (at a rest-frame
wavelength $\lambda_0 \sim 3950$\AA), which is the spectral signature of elliptical galaxies, plus
other absorption lines such as the {\it G} band ($\lambda_0 = 4300$\AA), H$\beta$ ($\lambda_0 =
4861$\AA), and the Mg {\sc II} triplet ($\lambda_0 \sim 5175$\AA), plus the [O {\sc II}] emission
line at rest-frame $\lambda_0 = 3727$\AA. The Na {\sc I} absorption doublet ($\lambda_0 \sim
5892$\AA) is also observable in the low-$z$ clusters.

\subsubsection{VLT-FORS2 Observations}

The FORS2-2009B observations (Run 1) were aimed at newly SZE-detected clusters regarded as
``secure'' candidates detected with ACT in 2008. These clusters had already been reported as SZE
detections by \cite{Staniszewski-SPT} and their physical properties characterized in
\cite{Menanteau-09}.

Run 3 was mostly focused on getting detailed information for ACT-CL~J0102$-$4915 \citep[``El
Gordo,''][]{Menanteau-J0102}, which was detected as the largest decrement in the ACT maps.
ACT-CL~J0559$-$5249 was also included in this run. 

Runs 1 and 3 were executed in Service Mode in semesters 2009B and 2010B, respectively. The
instrument setup in both runs was the same, using the GRIS 300I+11 grism and $1\arcsec$--wide slits,
which provides a resolving power $R=660$ at $\lambda=8600$\AA. A total of 18 FORS2/MXU masks were
observed for the five clusters. Each mask was observed for 40 minutes, which we estimated to be the
best compromise between maximizing S/N and number of masks.

FORS2 has a field of view of $6.\!\arcmin8\times6.\!\arcmin8$ in the standard resolution setup,
which corresponds to a width of $2517\,\,h_{70}^{-1}\,\rm{kpc}$ at $z=0.5$.

\subsubsection{Gemini-GMOS Observations}

The GMOS-2009B observations (Run 2) were aimed at four optically selected clusters from the SCS
whose richness-based mass estimates suggested that they would be detected by ACT in the SZE survey
\citep{Menanteau-10}. However, as mentioned above, only one object was in fact detected by ACT
(ACT-CL~J0521$-$5104); the other three are discussed in Section \ref{sec:SCS}. The total integration
time per mask was 3600~s ($2\times1800$~s). Two exposures at slightly different central wavelengths
per mask were required to cover the two 37-pixel gaps between the CCDs which run across the
dispersion axis.

Targets for Run 4 (GMOS-2010B) were selected from the sample of clusters newly discovered by ACT
presented in \cite{Marriage-ACT} and optically confirmed by \cite{Menanteau-ACT}. Run 4 was the only
one executed in Classical Mode, during five consecutive nights (December 6--10), all with clear,
photometric conditions and seeing $\lesssim0.\!\arcsec8$. Based on our experience in Run 2 we
decided to reduce the integration time to 2400~s ($2\times1200$s) for each mask during Run 4. This,
coupled with a $\sim20\%$ higher efficiency than Queue Mode, allowed us to observe a larger number
of masks (and clusters) while still obtaining the necessary S/N in the relevant spectral lines.

In both GMOS runs we used the R400\_G5325 grating and $1\arcsec$-wide slits, providing a resolving
power of $R\sim800$ with a $2\times2$ binning at $\lambda\sim7000$\AA. In the standard setup GMOS
has a field of view of $5.\!\arcmin5\times5.\!\arcmin5$ ($2036\,\,h_{70}^{-1}\,\rm{kpc}$ at
$z=0.5$).

\subsubsection{Data Reduction}

We have developed reduction pipelines both for the FORS2 and GMOS data, based on the existing
software by ESO and Gemini respectively, which work with IRAF/PyRAF\footnote{The pipeline used to
reduce GMOS data---dubbed ``\texttt{pygmos}''---is available at
\url{http://www.strw.leidenuniv.nl/\~sifon/pygmos/}.}. Cosmic rays are removed using L.A.Cosmic
\citep{LACos} with a detection limit of $4.5\sigma$. The wavelength calibrations were done using
CuAr lamps in the case of GMOS data and HeAr lamps for VLT data. The sky is subtracted from each
spectrum using a constant value determined locally within each slitlet.  In the case of GMOS data,
the individual exposures are coadded at this point. Finally, the one-dimensional (1D) spectra are
extracted from each slit and matched with the input photometric catalogs used to generate the masks.

\section{Analysis and Results}

\subsection{Galaxy Redshifts}

Galaxy redshifts are measured by cross-correlating the spectra with galaxy spectral templates of the
SDSS DR7 using the RVSAO/XCSAO package for IRAF \citep{RVSAO}; the spectral features in each
spectrum have been confirmed with the 2D spectra by visual inspection. We have been able to estimate
reliable redshifts for $\sim1200$ galaxies which comprise $\sim80\%$ of all targeted objects.

The median rms in the wavelength calibration is $\sim0.3$\AA\ and is similar for both instruments.
At a central wavelength of 6000~\AA, this corresponds to a velocity uncertainty of
$15\,\mathrm{km\,s^{-1}}$, which is within the errors of the cross-correlation velocity. In
particular, the latter is typically $\Delta(cz)\sim40-80\,\mathrm{km\,s^{-1}}$, as calculated by
RVSAO. It has been established experimentally that the true cross-correlation errors are larger than
those reported by RVSAO, by a factor $\sim1.7$ \citep[e.g.,][]{Quintana-00}, strengthening the point
that the calibration errors are well within the velocity measurement errors.

We have included the member catalog for ACT-CL~J0546$-$5345 published by \cite{Brodwin-SPT}. Seven
galaxies have been observed both by \cite{Brodwin-SPT} and by us; all redshifts are consistent
within $2\sigma$. We therefore use our measurements for those galaxies in the following analysis.

\subsection{Cluster Redshifts and Velocity Dispersions}\label{sec:vdisp}

%%%%%%%%%%%%%%%%%%%%%%%%%%%%%%%%%
%%%  Velocity Profile Figure  %%%
%%%%%%%%%%%%%%%%%%%%%%%%%%%%%%%%%
\begin{figure}
\centerline{\includegraphics[width=3.55in]{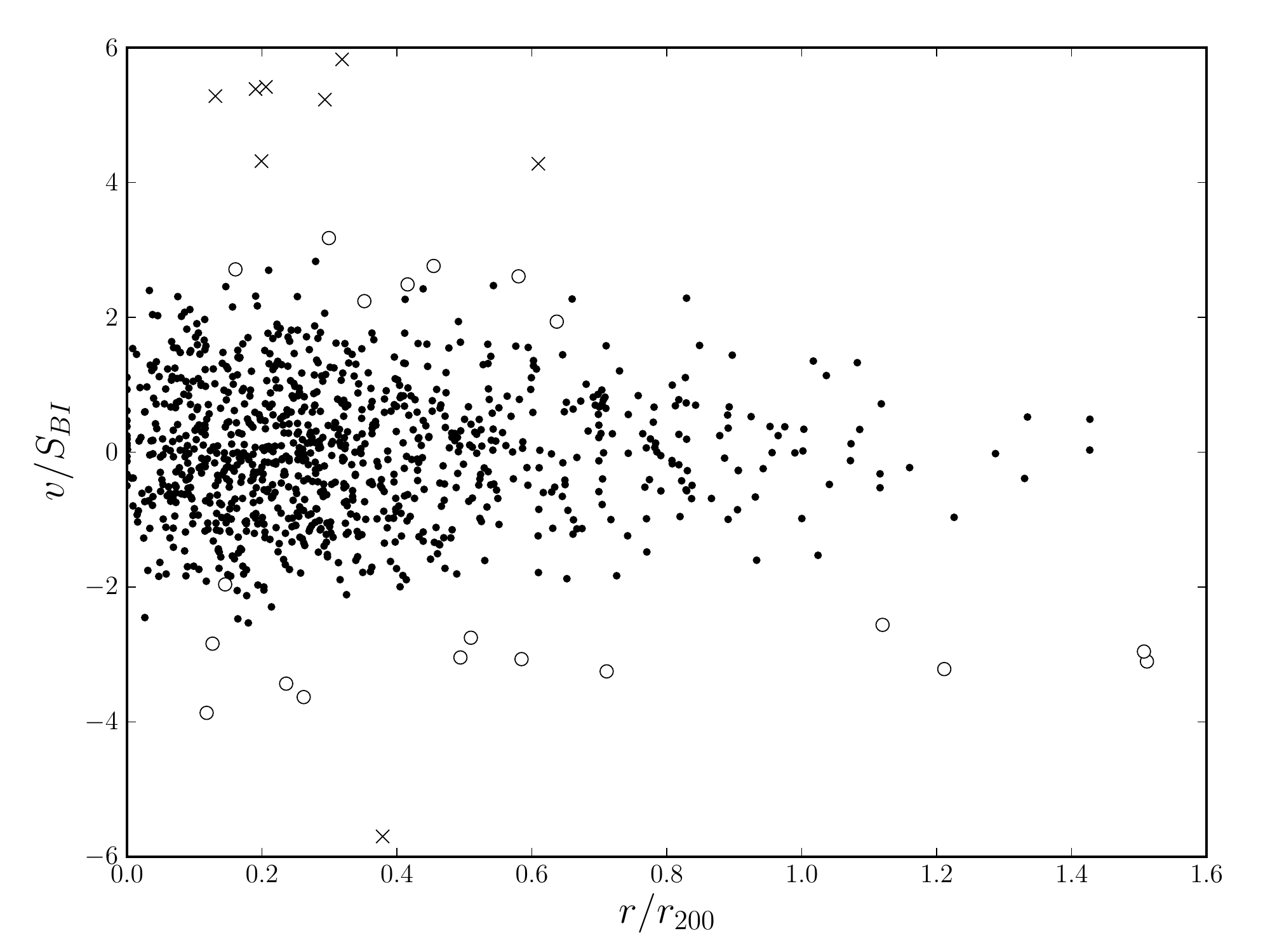}}
\caption{``Stacked'' result of the shifting gapper method of member selection, showing the
         galaxies in all 16 SZE-detected clusters. The horizontal axis shows the cluster-centric
         distance normalized by $r_{200c}$ for each cluster and the vertical axis shows the peculiar
         velocity of each galaxy, normalized by the velocity dispersion of the corresponding  
         cluster. Black dots show member galaxies, open circles show galaxies rejected by the
         method, and crosses show galaxies with peculiar velocities larger than
         $4000\,\mathrm{km\,s^{-1}}$. Galaxies with peculiar velocities larger than $6S_{\rm BI}$
         are not shown for clarity.}
\label{fig:rv}
\end{figure}

%%%%%%%%%%%%%%%%%%%%%%%%%%%%%%%%
%%%     Redshifts Figure     %%%
%%%%%%%%%%%%%%%%%%%%%%%%%%%%%%%%
\begin{figure}
\centerline{\includegraphics[width=3.55in]{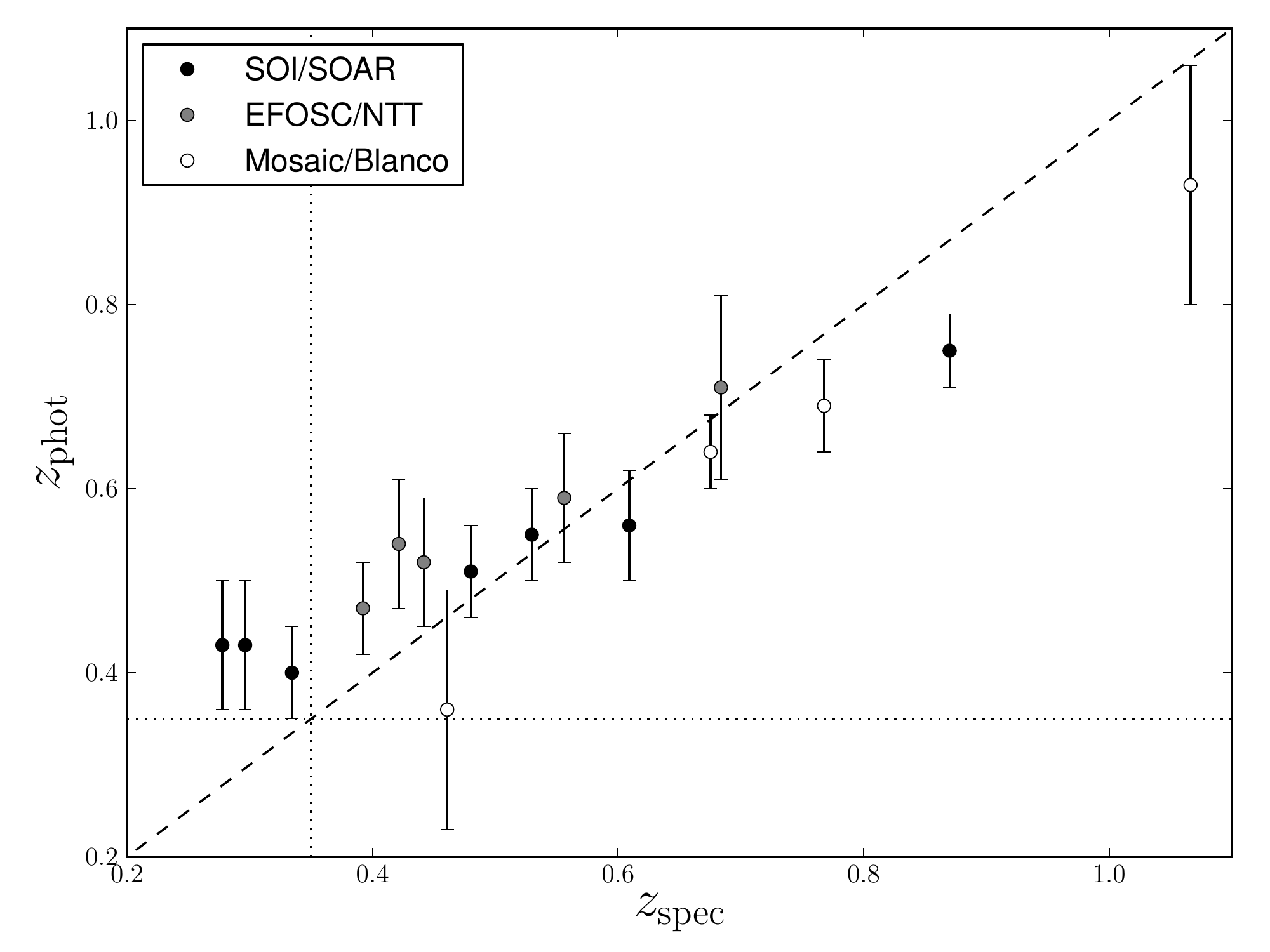}}
\caption{Comparison between spectroscopic redshifts from this work with initial $gri$ photometric
         redshift estimates from \cite{Menanteau-ACT}. The instrument and telescope with which each
         cluster was observed are identified in the legend. The dashed line shows $z_{\rm phot} =
         z_{\rm spec}$. The dotted horizontal line shows the sample selection cut, $z_{\rm phot} =
         0.35$, and the dotted vertical line shows the corresponding $z_{\rm spec} = 0.35$.}
\label{fig:spec_z}
\end{figure}

It is of great importance to correctly determine cluster membership to avoid a biased measurement of
the velocity dispersion \citep{Beers-91}. This is a complicated problem and many methods have been
developed to handle it. In this analysis, membership of galaxies to a cluster is determined by
applying a cut in (rest frame) velocity space of $4000\,\mathrm{km\,s^{-1}}$, and then applying the
shifting gapper method \citep{Fadda-96}. To do this, we define annular bins around the BCG, each of
which has at least 15 galaxies and radial width $\ge 250\,h_{70}^{-1}\,{\rm kpc}$. We consider the
histogram of velocities of member galaxies within each bin. We assume the profile is symmetric, and
identify the main body of galaxies as those whose velocity is bounded by gaps of $\ge
500\,\,\mathrm{km\,s^{-1}}$. Following \cite{Katgert-96} and \cite{Fadda-96}, galaxies separated
from the main body by $\ge1000\,\,\mathrm{km\,s^{-1}}$ are considered interlopers and are removed.
The selection method is iterated until the number of members is stable. This usually happens after
the second iteration. A total of 948 galaxies ($\sim65\%$ of all targets) have been selected as
cluster members. Most of these galaxies show the spectral signatures of elliptical galaxies and do
not have emission lines, and only a few emission-line galaxies belong to clusters (see
Section \ref{sec:color}). The galaxies remaining at this point are considered members of the
cluster. Figure \ref{fig:rv} shows the ``stacked'' result of this method, with members as solid
dots. The values have been normalized to allow for direct comparison of all clusters. We have
explored systematic effects coming from the member selection method by changing the width of the
bins, the number of galaxies per bin, and the size of either gap in the shifting gapper. Varying
these parameters yields results that are consistent with the reported velocity dispersions.

We use the biweight estimators of location \citep[hereafter $z_{\rm BI}$;][]{Beers-90} for the
redshift of the cluster and scale, $S_{\rm BI}$, for the velocity dispersion. All errors have been
estimated with the bootstrap resampling technique with 5000 iterations. The redshifts of the
clusters are presented in Figure \ref{fig:spec_z}, where they are compared to the photometric
redshifts of \cite{Menanteau-ACT}. The median redshift of the sample is $z=0.50$. The slightly
biased photometric redshifts apparent in Figure \ref{fig:spec_z} are mainly due to two factors: the
lack of a well-characterized filter response function for the telescopes involved in the imaging
follow-up and the use of only three to four filters for the determination of photometric redshifts
\citep{Menanteau-ACT}.

\cite{Danese-80} showed that the observational errors on the redshifts of galaxies introduce a bias
in the measured velocity dispersion. However, for a cluster of $M\sim10^{15}\,M_\odot$ with
individual errors as measured in this work (i.e., $\lesssim100\,\mathrm{km\,s^{-1}}$), this
correction is $<0.1\%$ (and even lower for more massive clusters), and it is therefore not
considered here\footnote{As mentioned before, the errors calculated by RVSAO are smaller than the
true cross-correlation errors. Even so, the \cite{Danese-80} correction would be $\ll1\%$, and
still negligible over the statistical uncertainty in the velocity dispersion.}.

\section{Dynamical Masses}
\label{sec:dyn_mass}

%%%%%%%%%%%%%%%%%%%%%%%%%%%
%%%  Dyn. Masses Table  %%%
%%%%%%%%%%%%%%%%%%%%%%%%%%%
\def\tnb{\tablenotemark{b}}
\def\tnc{\tablenotemark{c}}
\def\tnd{\tablenotemark{d}}
\def\tne{\tablenotemark{e}}
\def\tnf{\tablenotemark{f}}
\def\tng{\tablenotemark{g}}
\def\tnh{\tablenotemark{h}}
\def\tni{\tablenotemark{i}}
\begin{deluxetable*}{l c c c c c}
\tablecaption{Dynamical Properties of ACT 2008 Clusters}
\tablehead{
\colhead{ACT Descriptor} &
\colhead{$N_{\rm gal}$\tna} &
\colhead{$z_{\rm BI}$} &
\colhead{$S_{\rm BI}$} &
\colhead{$r_{200c}$} &
\colhead{$M_{200c}$}
 \\
\colhead{} &
\colhead{} &
\colhead{} &
\colhead{$(\mathrm{km\,s^{-1}})$} &
\colhead{$(h_{70}^{-1}~\rm{kpc})$} &
\colhead{$(10^{14}\,h_{70}^{-1}\,M_\odot)$}
}
\startdata
ACT-CL~J0102$-$4915\tnb & 89 & $0.8701\pm0.0009$ &    $1321\pm106$    & $1789\pm140$ &   
$16.3\pm3.8$ \\
ACT-CL~J0215$-$5212     & 55 & $0.4801\pm0.0009$ &    $1025\pm102$    & $1736\pm173$ &
\phn$9.6\pm2.8$ \\
ACT-CL~J0232$-$5257     & 64 & $0.5559\pm0.0007$ & \phn$884\pm110$    & $1438\pm177$ &
\phn$5.9\pm2.2$ \\
ACT-CL~J0235$-$5121     & 82 & $0.2777\pm0.0005$ &    $1063\pm101$    & $2007\pm190$ &   
$11.9\pm3.4$ \\
ACT-CL~J0237$-$4939     & 65 & $0.3344\pm0.0007$ &    $1280\pm89$\phn & $2339\pm162$ &   
$20.0\pm4.2$ \\
ACT-CL~J0304$-$4921     & 71 & $0.3922\pm0.0007$ &    $1109\pm89$\phn & $1971\pm155$ &   
$12.7\pm3.0$ \\
ACT-CL~J0330$-$5227\tnc & 71 & $0.4417\pm0.0008$ &    $1238\pm98$\phn & $2138\pm166$ &   
$17.1\pm4.0$ \\
ACT-CL~J0346$-$5438     & 88 & $0.5297\pm0.0007$ &    $1075\pm74$\phn & $1770\pm122$ &   
$10.7\pm2.2$ \\
ACT-CL~J0438$-$5419\tnd & 65 & $0.4214\pm0.0009$ &    $1324\pm105$    & $2310\pm182$ &   
$21.1\pm5.0$ \\
ACT-CL~J0509$-$5341\tne & 76 & $0.4607\pm0.0005$ & \phn$846\pm111$    & $1451\pm189$ &
\phn$5.5\pm2.1$ \\
ACT-CL~J0521$-$5104\tnf & 24 & $0.6755\pm0.0016$ &    $1150\pm163$    & $1744\pm245$ &   
$12.1\pm5.1$ \\
ACT-CL~J0528$-$5259\tng & 55 & $0.7678\pm0.0007$ & \phn$928\pm111$    & $1337\pm159$ &
\phn$6.1\pm2.2$ \\
ACT-CL~J0546$-$5345\tnh & 48 & $1.0663\pm0.0014$ &    $1082\pm187$    & $1319\pm226$ &
\phn$8.1\pm4.2$ \\
ACT-CL~J0559$-$5249\tni & 31 & $0.6091\pm0.0014$ &    $1219\pm118$    & $1916\pm184$ &   
$14.9\pm4.3$ \\
ACT-CL~J0616$-$5227     & 18 & $0.6838\pm0.0019$ &    $1124\pm165$    & $1699\pm244$ &   
$11.2\pm4.9$ \\
ACT-CL~J0707$-$5522     & 58 & $0.2962\pm0.0005$ & \phn$832\pm82$\phn & $1561\pm156$ &
\phn$5.7\pm1.7$ 
\enddata
\tablenotetext{a}{Number of spectroscopically confirmed members, after applying the selection
                  procedure of Section \ref{sec:vdisp}.}
\tablenotetext{b}{``El Gordo'' \citep{Menanteau-J0102} and SPT-CL~J0102$-$4915
                  \citep{Williamson-11}.}
\tablenotetext{c}{A3128 (NE) \citep{Werner-07}.}
\tablenotetext{d}{PLCK~G262.7$-$40.9 \citep{Planck-clusters} and SPT-CL~J0438$-$5419
                  \citep{Williamson-11}.}
\tablenotetext{e}{SPT-CL~J0509$-$5341 \citep{Staniszewski-SPT}.}
\tablenotetext{f}{SCSO~J052113$-$510418 \citep{Menanteau-10} and SPT-CL~J0521$-$5104
                  \citep{Vanderlinde-SPT}.}
\tablenotetext{g}{SPT-CL~J0528$-$5259 \citep{Staniszewski-SPT} and SCSO~J052803$-$525945
                  \citep{Menanteau-10}.}
\tablenotetext{h}{SPT-CL~J0547$-$5345 \citep{Staniszewski-SPT}.}
\tablenotetext{i}{SPT-CL~J0559$-$5249 \citep{Vanderlinde-SPT}.}
\label{t:dyn}
\end{deluxetable*}

In this section, we use the velocity dispersions measured in the previous section to estimate
cluster masses. The dynamical state of each cluster is also studied, including signs of substructure
and the fraction and influence of emission-line galaxies in the cluster population. Both factors
can, in principle, bias the velocity dispersion and thus the dynamical mass of the cluster.
Moreover, they are not expected to be completely independent, since emission-line galaxies are
generally newly incorporated galaxies, which might mean recent (or near-future) mergers involving
the main cluster \citep{Moore-99,Book-10}.

\subsection{Dynamical Mass Estimates}\label{sec:masses}

The relationship between velocity dispersions and masses has been the focus of several studies. As a
first-order approach, \cite{Heisler-85} studied simple variations of the Virial Theorem and found
that they all behave similarly, and that it is not possible to distinguish among them.
\cite{Carlberg-97} compared masses obtained from the Virial Theorem to those obtained with the Jeans
equation in observed clusters. They found that the former are biased high by a factor of 10\%--20\%
and associated this bias with a surface pressure correction factor of the same order.

More recently and based on large cosmological simulations, \cite{Evrard-08} concluded that massive
($M_{200c} > 10^{14}M_\odot$) clusters are, on average, consistent with a virialized state, and find
a best-fit scaling relation for dark matter halos described by NFW profiles in a variety of
cosmologies. Accordingly, the mass enclosed within $r_{200c}$ is

\begin{equation}\label{eq:mass}
 M_{200c} = \frac{10^{15}}{0.7h_{70}(z)} \left(\frac{\sigma_{\rm
   DM}}{\sigma_{15}}\right)^{1/\alpha}M_\odot\,\,,
\end{equation}

\noindent where $\sigma_{15} = 1082.9\pm4.0\,\,\mathrm{km\,s^{-1}}$, $\alpha =
0.3361\pm0.0026$,\break $h_{70}(z) = h_{70}\sqrt{\Omega_\Lambda + (1 + z)^3\Omega_M}$ for a flat
cosmology and $\sigma_{\rm DM}$ is the 1D velocity dispersion of the dark matter particles within
$r_{200c}$, which is related to the velocity dispersion of galaxies by a so-called bias factor $b_v
= S_{\rm BI}/\sigma_{\rm DM}$. As summarized by \cite{Evrard-08}, the bias factor as currently
estimated is $\langle b_v\rangle = 1.00 \pm 0.05$. For consistency with previous studies
\citep[e.g.,][]{Brodwin-SPT}, we adopt a value $b_v=1$, meaning that galaxies are unbiased tracers
of the mass in a cluster.

The mass values drawn from Equation (\ref{eq:mass}) are shown in Table \ref{t:dyn}, and the given
errors include uncertainties on the cluster redshift, the velocity dispersion, $\alpha$ and
$\sigma_{15}$. The overall uncertainty in the mass is dominated by statistical errors which, in
turn, are dominated by the error in the velocity dispersion. The systematics introduced by
Equation (\ref{eq:mass}) contribute $<10\%$ of the uncertainties listed in Table \ref{t:dyn}. The
mass from Equation (\ref{eq:mass}) yields a lower value than the virial mass estimator, as
\cite{Carlberg-97} also anticipated.

As indicated by \cite{Evrard-08}, Equation (\ref{eq:mass}) holds for primary halos, i.e., clusters
where a ``main system'' can be easily identified and substructure is only marginal. As noted in
Section \ref{sec:sub}, a high fraction of the clusters have significant substructure, but none of
them shows a clear bimodal distribution in velocity and we therefore assume that Equation
(\ref{eq:mass}) is applicable to all the clusters in the sample.

The radius $r_{200c}$ is also listed for each cluster in Table \ref{t:dyn}. These have been
calculated using $M_{200c}$ and assuming spherical clusters (i.e., $M_{200c} = 200\rho_c\times4\pi
r_{200c}^3/3$).

\subsection{Substructure}\label{sec:sub}

%%%%%%%%%%%%%%%%%%%%%%%%%%%
%%% Substructure Table  %%%
%%%%%%%%%%%%%%%%%%%%%%%%%%%
\begin{deluxetable*}{l c c c c c c c c}
\tablecaption{Substructure in ACT 2008 Clusters}
\tablehead{
\colhead{Cluster} &
\colhead{$z$} &
\colhead{$\lvert v_{\rm pec}\rvert$\tna} &
\colhead{$\lvert v_{\rm pec}\rvert/S_{\rm BI}$} &
\colhead{$\Delta r$\tnb} &
\colhead{$\Delta r/r_{200c}$} &
\colhead{s.l.~(DS)\tnc} &
\multicolumn{2}{c}{Disturbed?\tnd}
 \\
\colhead{} &
\colhead{} &
\colhead{$(\mathrm{km\,s^{-1}})$} &
\colhead{} &
\colhead{$({\rm arcsec})$} &
\colhead{} &
\colhead{} &
\colhead{} &
\colhead{}
}
\startdata
ACT-CL~J0102$-$4915\tne & $0.870$ & \phn\phn$10\pm169$ & $0.01\pm0.13$ &    \phn68 & 0.30 &
$0.48_{-0.02}^{+0.13}$ & 010 & Yes \\[0.1cm]
ACT-CL~J0215$-$5212     & $0.480$ &       $1171\pm153$ & $1.14\pm0.19$ &    \phn33 & 0.12 &
$0.02_{-0.01}^{+0.00}$ & 101 & Yes \\[0.1cm]
ACT-CL~J0232$-$5257     & $0.556$ & \phn\phn$37\pm129$ & $0.04\pm0.14$ &    \phn35 & 0.15 &
$0.11_{-0.05}^{+0.11}$ & 000 & No \\[0.1cm]
ACT-CL~J0235$-$5121     & $0.278$ &    \phn$138\pm137$ & $0.13\pm0.13$ &    \phn44 & 0.09 &
$0.04_{-0.03}^{+0.01}$ & 001 & Yes \\[0.1cm]
ACT-CL~J0237$-$4939     & $0.334$ &    \phn$261\pm174$ & $0.20\pm0.14$ &    \phn78 & 0.16 &        
$<0.01$        & 001 & Yes \\[0.1cm]
ACT-CL~J0304$-$4921     & $0.392$ &    \phn$151\pm157$ & $0.14\pm0.14$ &    \phn22 & 0.06 &
$0.04_{-0.03}^{+0.09}$ & 001 & No \\[0.1cm]
ACT-CL~J0330$-$5227     & $0.442$ &    \phn$424\pm167$ & $0.34\pm0.14$ &    \phn44 & 0.12 &
$0.21_{-0.02}^{+0.27}$ & 100 & No \\[0.1cm]
ACT-CL~J0346$-$5438     & $0.530$ &    \phn$263\pm125$ & $0.24\pm0.12$ &    \phn16 & 0.06 &
$0.23_{-0.07}^{+0.05}$ & 100 & No \\[0.1cm]
ACT-CL~J0438$-$5419     & $0.421$ &    \phn$392\pm172$ & $0.30\pm0.13$ &    \phn10 & 0.02 &
$0.03_{-0.02}^{+0.01}$ & 101 & Yes \\[0.1cm]
ACT-CL~J0509$-$5341     & $0.461$ &    \phn$361\pm134$ & $0.42\pm0.17$ &       114 & 0.46 &
$0.08_{-0.03}^{+0.04}$ & 110 & Yes \\[0.1cm]
ACT-CL~J0521$-$5104\tnf & $0.676$ &    \phn$440\pm292$ & $0.37\pm0.25$ &    \phn37 & 0.15 &         
\ldots            & 00- & No? \\[0.1cm]
ACT-CL~J0528$-$5259     & $0.768$ &    \phn$144\pm177$ & $0.16\pm0.19$ &    \phn50 & 0.28 &
$0.30_{-0.02}^{+0.07}$ & 010 & No \\[0.1cm]
ACT-CL~J0546$-$5345     & $1.066$ &    \phn$541\pm163$ & $0.50\pm0.17$ &    \phn20 & 0.13 &
$0.02_{-0.02}^{+0.04}$ & 101 & Yes \\[0.1cm]
ACT-CL~J0559$-$5249     & $0.609$ &    \phn$233\pm241$ & $0.19\pm0.20$ & \phn\phn9 & 0.03 &
$0.13_{-0.06}^{+0.13}$ & 000 & No \\[0.1cm]
ACT-CL~J0616$-$5227\tnf & $0.684$ &    \phn$685\pm268$ & $0.61\pm0.25$ &    \phn29 & 0.12 &         
\ldots           & 10- & Yes? \\[0.1cm]
ACT-CL~J0707$-$5522     & $0.296$ &    \phn$402\pm140$ & $0.48\pm0.18$ &    \phn19 & 0.05 &
$0.34_{-0.15}^{+0.04}$ & 100 & No
\enddata
\tablecomments{Redshifts are listed for reference.}
\tablenotetext{a}{Absolute value of the peculiar velocity of the BCG in the cluster rest-frame (see
                  Section \ref{sec:vpec}). The uncertainties consider the error on the BCG redshift
                  as twice that given by RVSAO.}
\tablenotetext{b}{Offset between the BCG and the SZ peak as found in the $Y_{200c}$ analysis (see 
                  Sections \ref{sec:sze} and \ref{sec:offset} for details).}
\tablenotetext{c}{Significance level of the DS test. Uncertainties are computed at the 75\% level 
                  (see Section \ref{sec:ds} for details).}
\tablenotetext{d}{Each ordered number represents one of the tests listed in the table: ``1'' means 
                  the test shows evidence for substructure and ``0'' means it does not.}
\tablenotetext{e}{This cluster is classified as ``disturbed'' based on the results of 
                  \cite{Menanteau-J0102}. See the text for details.}
\tablenotetext{f}{There are too few members observed for the DS test to be reliable. The 
                  classification is left as a tentative one, and these clusters are excluded from
                  the analysis of Section \ref{sec:bias} (see Section \ref{sec:subresults}).}
\label{t:sub}
\end{deluxetable*}

It is becoming widely accepted that substructure is a common feature of galaxy clusters, and that
its presence (or lack thereof) is related to the degree of relaxation and hence the validity of the
hydrostatic equilibrium hypothesis \citep[e.g.,][and references therein]{Battaglia-12}. While X-ray
observations can reveal the presence of substructure in the plane of the sky, velocity information
can reveal substructure in the radial direction. From X-ray observations over a wide range in masses
at $z\lesssim0.3$, \cite{Schuecker-01} find that $(52\pm7)\%$ of galaxy clusters present significant
substructure. \cite{Girardi-97} find that out of 44 optically selected local ($z\le0.15$) clusters,
15 (38\%) show significant signs of substructure based on their dynamics. \cite{Girardi-97} argue,
on the other hand, that substructure found in clusters that show a {\it unimodal} velocity
distribution (i.e., where the substructure is not of comparable size to the cluster itself) does not
influence the velocity dispersion (hence mass) measurements.

In general, a non-negligible fraction of the galaxy clusters in a sample will have biased mass
measurements due to substructure. These results highlight the need for a correct estimation of the
degree to which galaxy clusters seem to be relaxed or in the process of merging.

One very basic test for substructure involves the distribution of measured velocities.  In fact,
however, none of our velocity histograms shows clear evidence for a bi- or multi-modal distribution
and the velocity dispersions $S_{\rm BI}$ are consistent with Gaussian velocity dispersions (i.e.,
with the standard deviation), in all cases, within 1$\sigma$. So, in the following, we employ three
specific tests that take advantage of the three-dimensional (3D) information provided by the optical
spectroscopy to assess the dynamical state of the clusters from a wide perspective. Table
\ref{t:sub} summarizes the substructure analysis.

\subsubsection{1D: BCG Peculiar Velocity}\label{sec:vpec}

For a cluster that is relaxed, the peculiar velocity of the BCG should be close to zero \citep[][but
see \cite{Pimbblet-06} for a likely counter example]{Quintana-82,Oegerle-01}. \cite{Oegerle-01} find
that the dispersion of BCG peculiar velocities is $\sim160\,\mathrm{km\,s^{-1}}$ for a median
$S_{\rm BI}\sim800\,\mathrm{km\,s^{-1}}$. Using a sample of 452 Abell clusters, \cite{Coziol-09}
find that BCGs have a median peculiar velocity $0.32S_{\rm BI}$ and that 41\% of BCGs have
velocities different from zero at the $2\sigma$ level, but note that this number is comparable to
the fraction of clusters that show signs of substructure. In summary, velocities consistent with
zero are not necessarily expected. Dominant (D/cD) BCGs, however, are mostly found in the low
peculiar velocity regime. Thus here clusters are (provisionally) considered as disturbed if their
BCG has a peculiar velocity different from zero at the $2\sigma$-level where, following
\cite{Coziol-09}, the fractional uncertainties are given by

\begin{equation}
 \Delta\left(v_{\rm pec}/S_{\rm BI}\right) = \frac1{S_{\rm BI}}\sqrt{\left(\Delta v_{\rm
    pec}\right)^2 + \left(\frac{v_{\rm pec}\Delta S_{\rm BI}}{S_{\rm BI}}\right)^2}
\end{equation}

\noindent where $\Delta v_{\rm pec} = \sqrt{S_{\rm BI}^2/N_{\rm gal} + \left(\Delta v_{\rm
BCG}\right)^2}$ is the error in the peculiar velocity, and $\Delta v_{\rm BCG}$ is twice the
cross-correlation error estimated by RVSAO, which is a conservative correction \citep{Quintana-00}.

Eight clusters meet this criterion, which will be coupled with similarly chosen criteria in the 2D
and 3D analyses before selecting which clusters have significant evidence for substructure.

\subsubsection{2D: Projected BCG-SZE Offset}\label{sec:offset}

Under the hypothesis of hydrostatic equilibrium, galaxies closely trace the total mass distribution
in the cluster and thus the BCG is located at the peak of the gravitational potential. If the
cluster is virialized, the gas should also follow the mass distribution. Deviation from this
scenario may be quantified by an offset between the BCG (i.e., dark matter) and the SZE (i.e., gas)
peak. This, of course, is sensitive to offsets projected in the sky, unlike the preceding and
following tests.

ACT has a beam of $1.\!\arcmin4$ (FWHM) at 148~GHz \citep{Hincks-ACT} and the uncertainties in the
determination of the position of each cluster are of order 10$\arcsec$-15$\arcsec$. We therefore
list the projected offset in arcseconds in Table \ref{t:sub}; offsets $\lesssim15\arcsec$ are within
ACT's positional uncertainty and should therefore not be considered physical offsets. \cite{Lin-04}
find that $>80\%$ of BCGs are offset from the peak gas emission by $\Delta r/r_{200c}<0.2$.
Moreover, \cite{Skibba-11} find that $\sim40\%$ of BCGs do not sit at the minimum of the potential
well in clusters. Column 6 of Table \ref{t:sub} lists the projected offset between the BCG and the
SZE peak for each cluster relative to the characteristic scale of the cluster $r_{200c}$.

We choose $\Delta r/r_{200c}\sim0.20$ as the threshold between (tentatively classified) relaxed and
disturbed clusters, based on the results of \cite{Lin-04}. In this case, only three
clusters---ACT-CL~J0102$-$4915, ACT-CL~J0509$-$5341, and ACT-CL~J0528$-$5259---have values over the
threshold. Given that the chance of l.o.s. substructure should be the same as that of substructure
in the plane of the sky,\footnote{In fact, the latter should be approximately twice
as large, given the number of dimensions covered by the plane of the sky and the l.o.s..} this might
be too stringent a limit. Moreover, the findings of \cite{Skibba-11} argue that this might not be
a very reliable test for substructure, but we include it for completeness. The three clusters that
meet this criterion have offsets on the order of an arcminute, far beyond uncertainties in the ACT
SZE centroids and therefore qualify as physical offsets.

It is worth noting that positions estimated in our analysis differ from those reported in
\cite{Marriage-ACT}, typically by $\approx20\arcsec$. There are two exceptions, however: the
estimated centers for ACT-CL~J0509$-$5341 and ACT-CL~J0707$-$5522 have changed by 91\arcsec and
119\arcsec, respectively. These two clusters are also the clusters with the lowest S/N, as can be
seen from Table \ref{t:sze}, so these large shifts are attributed to this fact.

\subsubsection{3D: DS Test}\label{sec:ds}

By studying a large sample of statistical tests for substructure in galaxy clusters,
\cite{Pinkney-96} have shown that the DS test \citep{Dressler-88} is the most sensitive test when
used individually. The test has the ability not only to detect the presence of substructure, but
also to locate the latter in projected space (in the ideal cases of substructure not overlapping
with the main system neither in velocity nor in projected space) and is based in the detection of
localized subgroups of galaxies that deviate from the global distribution of velocities by use of
the parameter $\Delta = Section igma_i \delta_i$, where

\begin{equation}
 \delta_i^2 = \frac{N_{\rm local}}{\sigma^2}\left[\left(\bar v_i - \bar v\right)^2 + \left(\sigma_i 
     - \sigma\right)^2\right]^2
\end{equation}

\noindent is computed for each cluster member, where $\bar v_i$ and $\sigma_i$ are the mean and
standard deviation of the velocity distribution of the $N_{\rm local}$ members closest to the {\it
i}th member, and $\bar v$ and $\sigma$ are the mean and standard deviation of the velocity
distribution of all the cluster members. The significance level (s.l.)~of the test is obtained by
shuffling the velocities of each galaxy via a bootstrap resampling technique with 5000 iterations.
Although the common use is that $N_{\rm local}=\sqrt{N_{\rm gal}}$, in this work $\Delta$ is
calculated for $N_{\rm local}$ ranging from 5 to 12. The uncertainties in the s.l.~are given by the
second-maximum and second-minimum s.l.~for each cluster when varying $N_{\rm local}$ (i.e., they
correspond to $\sim75\%$-level uncertainties), and the central value is given by the median. A large
uncertainty (i.e., dependence on $N_{\rm local}$) might also be indicative of substructure, but we
do not include this in the analysis.

As shown by \cite{Pinkney-96}, the false positive rate for the DS test is $<1\%$, $<4\%$ and 9\% for
a s.l.~of 1\%, 5\% and 10\% respectively, for member samples as large as ours in clusters simulated
by Gaussian distributions of galaxies. The threshold for substructure detection is set therefore at
5\% s.l.~within uncertainties; seven clusters meet this criterion. Given a false positive rate of
$4\%$, there is a 25\% chance of a false detection of substructure by this method.

\subsubsection{Substructure Results}\label{sec:subresults}

Clusters have been identified as merging systems if they meet at least two of the three conditions
explained above, or if they have an s.l.~of the DS test {\it strictly} below 5\% within
uncertainties. Although the second of the three conditions depends on the projected spatial
distribution, it is clear that this analysis is biased toward l.o.s.~substructure.

ACT-CL J0102$-$4915 (``El Gordo'') is a special case, as it does not show evidence for merging from
the dynamical information alone. However, both the spatial galaxy distribution and X-ray surface
brightness distribution reveal that this is a very complex system where two massive clusters are
interacting close to the plane of the sky \citep{Menanteau-J0102}.

On the other hand, ACT-CL J0616$-$5227 is tentatively considered as a merging cluster given the high
peculiar velocity of the BCG, but the DS test was not performed for this cluster given the low
number of members. The latter note also applies to ACT-CL~J0521$-$5104, although this cluster is
tentatively considered relaxed. These two clusters have been excluded from the analysis of
Section \ref{sec:bias}.

The last column of Table \ref{t:sub} states whether a cluster is considered to be relaxed (``No'')
or disturbed (``Yes'') while the previous column lists whether each cluster shows (``1'') or does
not show (``0'') signs of substructure in each of the tests, as defined above. Combining the
criteria used, 7 out of 14 clusters show signs of merger activity (or 8 of 16, if we include
ACT-CL~J0521$-$5104 and ACT-CL~J0616$-$5227). This number is consistent with previous optical and
X-ray studies of local clusters \citep[e.g.,][]{Girardi-97, Schuecker-01} and is also consistent
with the X-ray follow-up of SPT SZE-detected clusters by \cite{Andersson-SPT}. They find that 9 out
of 15 SZE-selected clusters show signs of substructure based purely on X-ray morphology.

\subsection{The Influence of Emission-line Galaxies}\label{sec:color}

%%%%%%%%%%%%%%%%%%%%%%%%%%%%%%%%
%%%   Blue Fraction Figure   %%%
%%%%%%%%%%%%%%%%%%%%%%%%%%%%%%%%
\begin{figure}
\centerline{\includegraphics[scale=0.48]{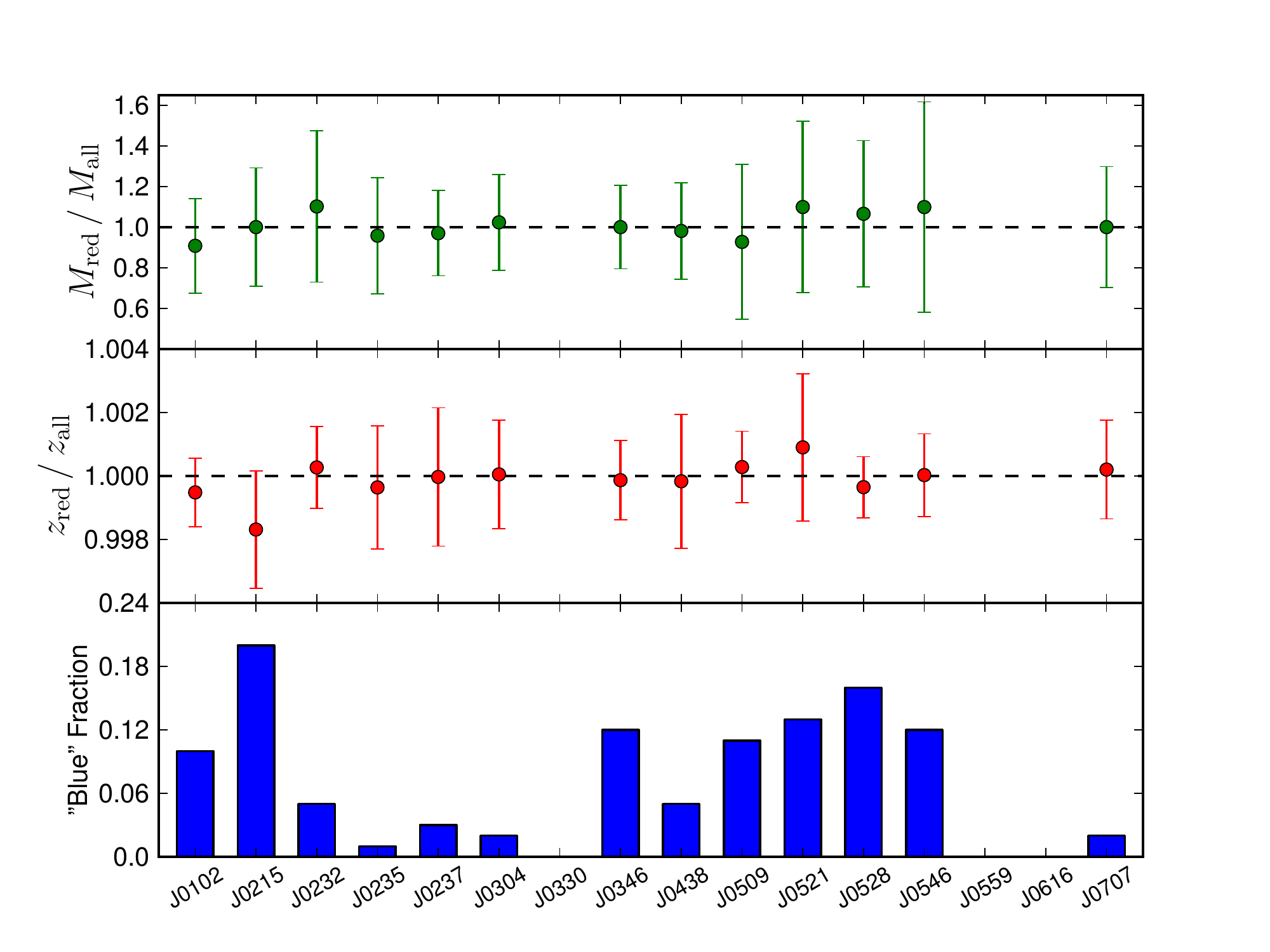}}
\caption{Top two panels show, for each cluster indicated on the horizontal axis, the ratios of
         dynamical masses {\it (top)} and cluster redshifts {\it (middle)} when only the
         absorption-line (``red'') galaxies or all galaxies are used for the analysis. Error bars
         are given by $\Delta M_{\rm all} / M_{\rm all}$ and $\Delta z_{\rm all} / z_{\rm all}$,
         respectively. The dashed line in each panel marks a ratio of unity. The bottom panel
         shows the observed fraction of galaxies with emission lines (``blue''). Cluster names have
         been shortened for clarity; data points in the top and middle panels have been omitted for
         the three clusters with blue fractions equal to 0.}
\label{fig:colorbias}
\end{figure}

Clusters of galaxies are mostly populated by passive galaxies. Late-type galaxies are preferentially
found in the outskirts of clusters and associated with infalling groups. They therefore tend to show
a different velocity distribution \citep{Biviano-04}. \cite{Girardi-96} find that 29\% (53\%) out of
a sample of 17 nearby clusters show differences in the velocity dispersion and 24\% (47\%) in the
mean velocity, at the $2\sigma$ ($1\sigma$) level. Simulations also show that, where blue galaxies
are found (i.e., outside the core), they tend to have a higher velocity dispersion than red galaxies
\citep{Springel-01b}. The way blue galaxies are distributed in the cluster (both in space and in
velocity) depends, however, on the history of each cluster \citep{Biviano-04}. The issue is complex;
for example, \cite{Aguerri-10} find no difference in the fraction of blue galaxies between relaxed
and disturbed clusters.

Although our target selection procedure should not be strongly biased against emission-line
galaxies, the observations have not been designed to study this effect and the spectroscopic samples
have emission-line fractions of $\lesssim10\%$ in most cases. This number does not necessarily
reflect the actual fraction in the clusters and could be taken as a lower limit for it. In spite of
all this, we briefly study the effect that blue\footnote{Although the classification is done purely
based on the spectral features of each galaxy (with or without emission lines), we sometimes speak
of blue and red, instead of emission- and absorption-line galaxies, respectively, for convenience.}
galaxies might have on the results.

Figure \ref{fig:colorbias} shows, in the top and middle panels respectively, the variation of the
mass measurement and the cluster redshift when blue (i.e., emission-line) galaxies are, and are not,
included. The null hypothesis (i.e., no bias) corresponds to $M_{\rm red}/M_{\rm all} = 1$.
Uncertainties in $z_{\rm red}/z_{\rm all}$ and $M_{\rm red}/M_{\rm all}$ are given by $\Delta z_{\rm
all}/z_{\rm all}$ and $\Delta M_{\rm all}/M_{\rm all}$, respectively, where $\Delta z_{\rm all}$ and
$\Delta M_{\rm all}$ are the uncertainties reported in Table \ref{t:dyn}. Within uncertainties,
neither cluster redshifts nor dynamical masses change when including, or not, emission-line
galaxies. Data points are not shown for the three clusters with fractions of emission-line
galaxies---which are shown in the bottom panel of Figure \ref{fig:colorbias}---equal to zero. These
three clusters have, by definition, $M_{\rm red}/M_{\rm all} = z_{\rm red}/z_{\rm all} = 1$. Note
that for the red-only analysis, the blue galaxies are removed before the selection process (i.e.,
$N_{\rm red}$ does not necessarily equal $N_{\rm gal} - N_{\rm blue}$).

The redshifts, velocity dispersions and corresponding masses in Table \ref{t:dyn} have been
calculated using all galaxies, since blue galaxies do not bias our mass (or redshift) measurements.
This is, in turn, consistent with the findings of \cite{Aguerri-10}.

\section{SZE--Mass Scaling Relations}\label{sec:scaling}

Both \cite{Vanderlinde-SPT} and \cite{Sehgal-ACT} have shown that, given an accurate calibration of
the SZE--mass scaling relation, the inclusion of the ACT or SPT cluster samples can lead to
significant improvements in cosmological parameter uncertainties, particularly $w$ and $\sigma_8$,
over \wmap-7 only constraints. These results have recently been confirmed by \cite{Benson-13} using
X-ray observations. However, without a precise SZE--mass scaling relation, these cluster samples do
not provide significant improvements in constraining cosmological parameters.

Observations have shown that the SZE signal and mass of a cluster can be related by a power-law
\citep{Benson-04, Bonamente-08, Melin-11}. While most simulations seem to confirm this
\citep{daSilva-04, Motl-05, Nagai-06}, others suggest that certain effects (e.g., AGN feedback) can
cause deviations from a single power-law dependence \citep{Battaglia-12}. In this work, we restrict
ourselves to a power-law relation between dynamical mass (see Section \ref{sec:dyn_mass} and Table
\ref{t:dyn}) and each SZE estimator measured from the ACT data (see Section \ref{sec:sze_par} and
Table
\ref{t:sze}) of this form:

\begin{subequations}\label{eq:scaling}
\begin{align}
 \frac{M_{200c}}{h_{70}^{-1}M_\odot} &=
  10^A\left(\frac{\widetilde{y_0}E(z)^{-2}}{5\times10^{-5}}\right)^B \label{eq:ytilde} \\[1ex]
 \frac{M_{200c}}{h_{70}^{-1}M_\odot} &=
  10^A\left(\frac{y_0E(z)^{-2}}{7\times10^{-5}}\right)^B \\[1ex]
 \frac{M_{200c}}{h_{70}^{-1}M_\odot} &= 
  10^A\left(\frac{Y_{200c}D_A(z)^2E(z)^{-2/3}}
                 {5\times10^{-5}\,h_{70}^{-2}\,\mathrm{Mpc^2}}\right)^B.
\end{align}
\end{subequations}

Here, $D_A(z)$ is the angular diameter distance in Mpc, $M_{200c}$ is in units of
$h_{70}^{-1}M_\odot$ and $E(z) = [\Omega_M(1 + z)^3 + \Omega_\Lambda]^{1/2}$. We refer to $B$ as the
(logarithmic) slope of the scaling relations. The self-similar predictions are 1 and 0.6 for the
$y_0$ and $Y_{200c}$ scaling relations, respectively \citep[e.g.,][]{Bonamente-08,Marriage-ACT}.
Equations (\ref{eq:scaling}) are convenient forms of parameterizing the scaling relations if one
wants to predict the mass of a cluster using SZE observations.

\subsection{Selection Biases}\label{sec:corrections}

Before proceeding, we consider the selection biases that can affect our study when fitting the
scaling relations  \citep[see][for a pedagogical description]{Mantz-10a}.

The first one is \cite{Eddington-13} bias, which results from the asymmetry of the steep underlying
mass function \citep[e.g.,][]{Jenkins-01,Tinker-08}, coupled with measurement errors, which
introduces a net shift in mass due to the statistical fluctuations of the measurement of the mass
proxy in the mass--observable (in this case, SZE signal) relation. While there are analytical
prescriptions to account for Eddington bias \citep[e.g.,][]{Mortonson-11}, we have assessed the
effect of measurement errors in our sample by simulating measurement uncertainties, comparable
to those of our data, in the simulations of \cite{Sehgal-10} and re-calculating the scaling relation
for 10,000 realizations. We find that the scaling relations in these simulations are unchanged when
introducing measurement uncertainties. We thus estimate that Eddington bias can be safely neglected
in this case.

The second bias is produced by the intrinsic scatter in the observable: clusters with mean SZE
fluxes at the detection limit whose signals scatter up will make it into the sample, while those
that scatter down will not. We refer to this effect as a flux bias.\footnote{This effect has often
been called ``Malmquist bias''. However, the term ``Malmquist bias'' was coined related to the
specific problem of an error in average distance modulus measurements tied with a magnitude-limited
sample and so is inappropriate here; see the review by \cite{Teerikorpi-97}.} We use the simulations
of \cite{Sehgal-10} including prescriptions for both AGN and supernova feedback, plus a realistic
modeling of non-thermal pressure support \citep{Bode-12}, to investigate this effect in our sample.
We measure $\widetilde{y_0}$ from the simulated clusters as in Section \ref{sec:sze_par} and
subsequently apply a cut $\widetilde{y_0}T_{\rm CMB}>150\mu K$ to the simulated data. This
``observational'' cut approximately reflects the detection threshold of the observed cluster sample.
This procedure mimics the observational situation with the exception that it assumes a constant
noise level throughout the survey. The ACT sample is defined in terms of an S/N limit, although the
noise level is approximately constant except near the edges of the map
\citep{Marriage-ACT,Marriage-ACTb}.  Within our sample, there is potentially only one cluster
(ACT-CL~J0707$-$5522) whose flux bias correction is not accurately described by this procedure
because it sits in a high-noise region in the maps. If this cluster is removed from the sample, the
change in the corrected scaling laws is negligible.

In practice, clusters within a mass range from $M$ to $M + \Delta M$ (where $\Delta
M=10^{14}\,h_{70}^{-1}M_\odot$) are extracted from the simulations and the average $\widetilde{y_0}$
value of the extracted subsample is determined both with and without a detection threshold. The
ratio of $\widetilde{y_0}$ values represents a statistical estimate of the flux bias factor for
clusters within this mass range. At the low mass end of the cluster sample the bias correction
factor is $\sim0.8$, while for clusters with $M_{200c}>9\times10^{14}M_{\odot}$ the correction
factor is close to unity. A continuous smooth curve is fitted to the bias correction factors as a
function of mass and applied individually to each cluster's SZE measurements. The uncertainty on the
mass is propagated through the bias correction factors and then into the corrected measurements. To
test the dependence of the correction on the adopted cluster physics, the procedure is repeated both
for a model with no star formation or AGN feedback and only thermal pressure support (an
``adiabatic'' model) and for a model with a generous 20\% non-thermal pressure support, constant
with mass, radius, and redshift \citep{Bode-09} as limiting cases, again accounting for the
uncertainty in the mass measurements. The variation in the scaling relations given by these
simulations is well within reported errors on the scaling relations, both in the normalizations and
in the slopes.

We apply the same bias correction factors to each of the different SZE estimators. This is a
reasonable approach since the latter are all based on matched filters with kernels of similar
scales. To distinguish the bias-corrected values hereafter, we label them with a superscript
``corr''.

\subsection{Best-fit Scaling Relations}\label{sec:fits}

%%%%%%%%%%%%%%%%%%%%%%%%%%%%%%%%%%
%%%  Scaling Relations Figure  %%%
%%%%%%%%%%%%%%%%%%%%%%%%%%%%%%%%%%
\begin{figure}
\centerline{\includegraphics[width=3.54in]{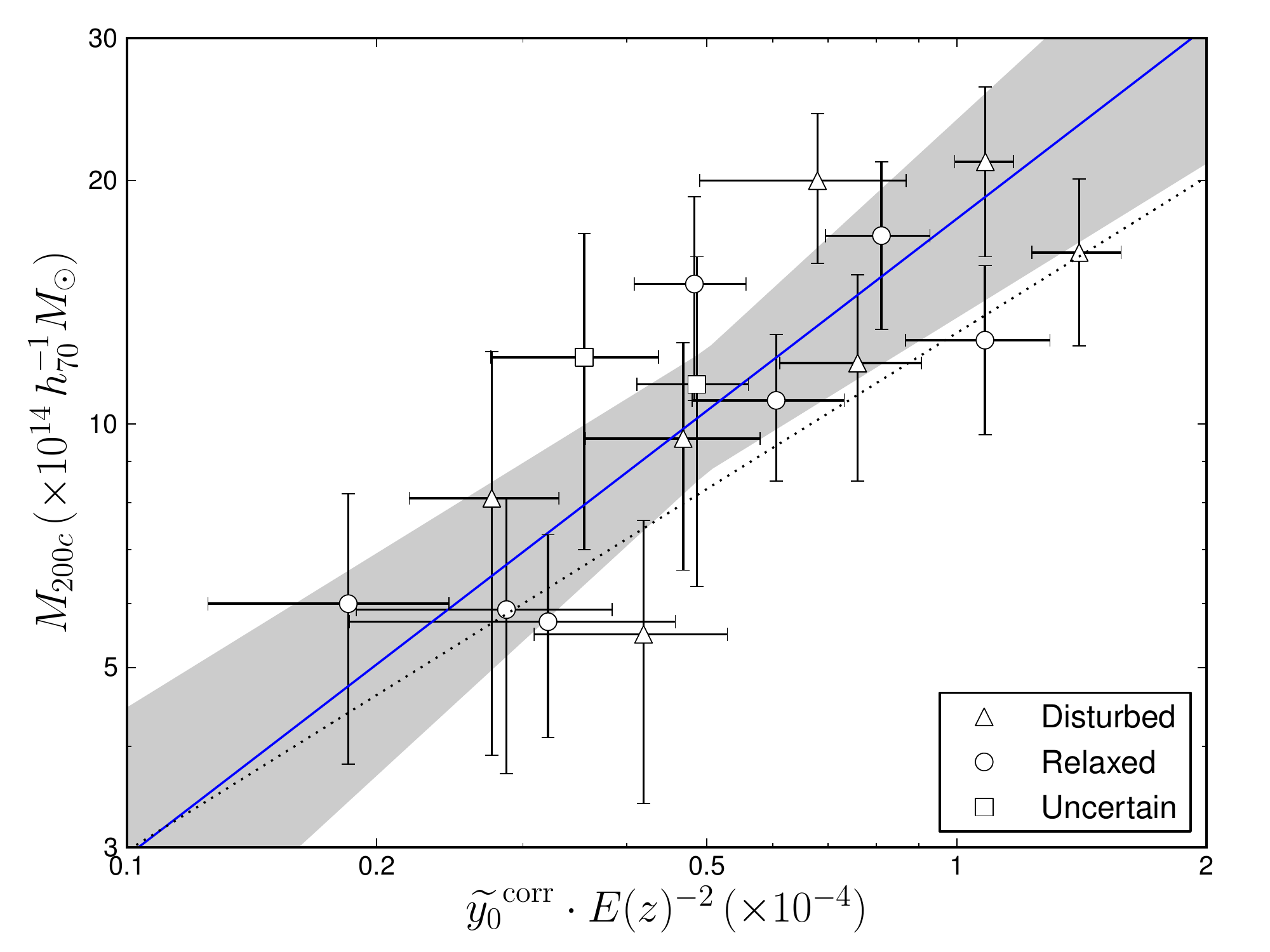}}
\centerline{\includegraphics[width=3.54in]{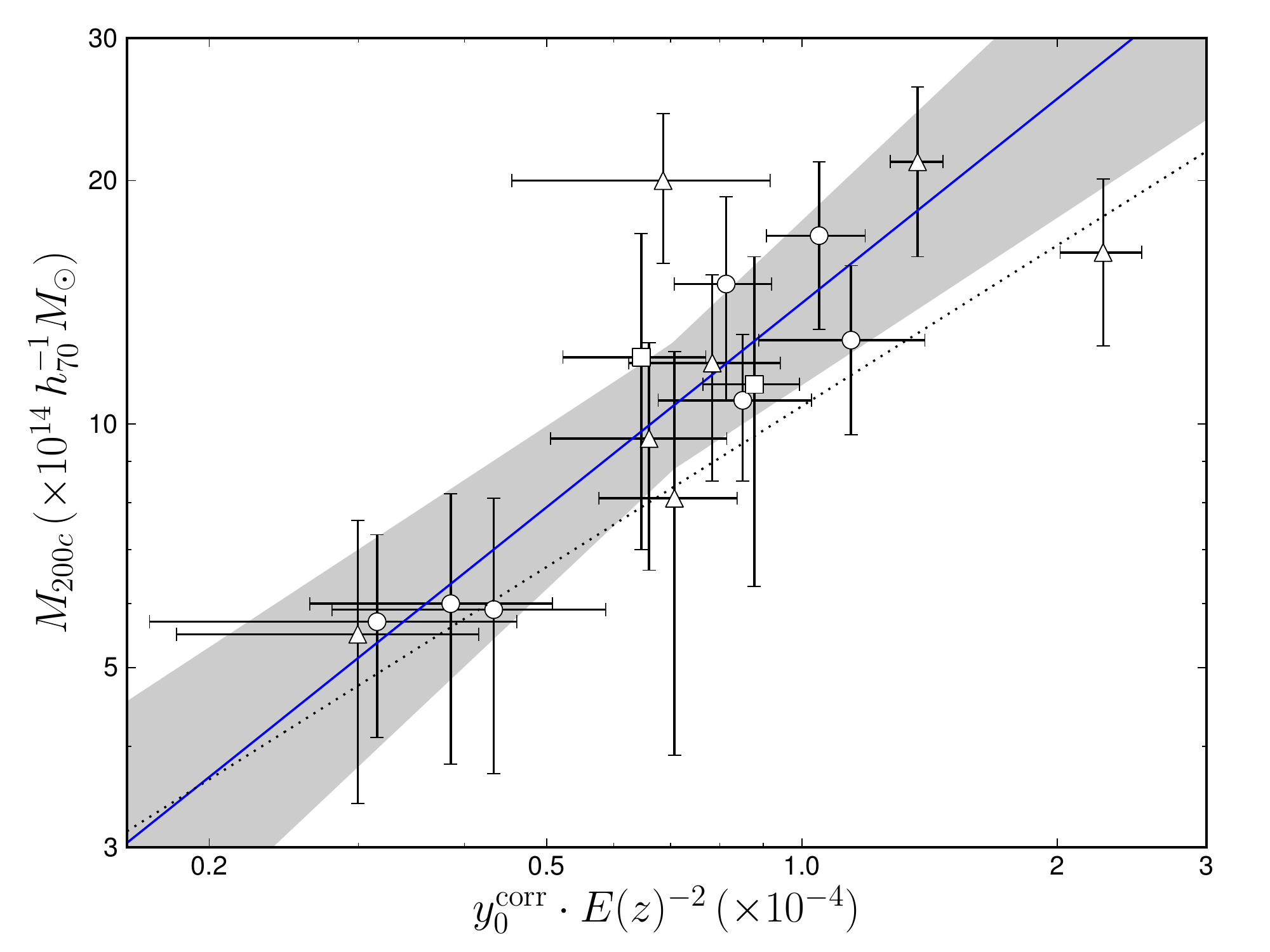}}
\centerline{\includegraphics[width=3.54in]{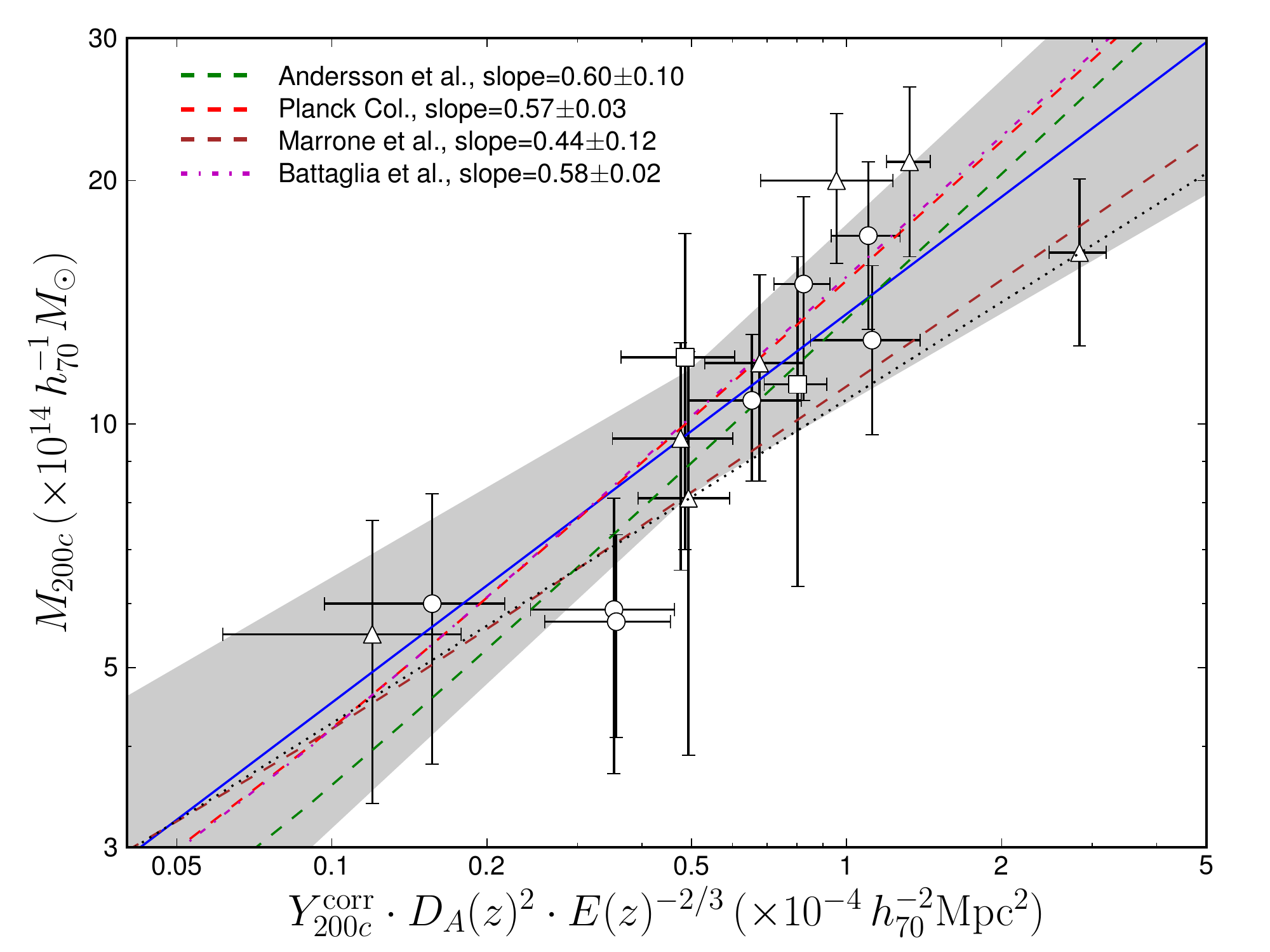}}
\caption{Scaling relations between SZE estimators and dynamical mass for the match-filtered
         amplitude $\widetilde{y_0}$ {\it (top)}, the central Compton parameter $y_0$ {\it (middle)}
         and $Y_{200c}$, the Compton $y$-parameter integrated out to $r_{200c}$ {\it (bottom)}, all
         including three-season ACT data. All estimators have been scaled as indicated in the axis
         labels (see Equations (\ref{eq:scaling})) and data points have been corrected for flux bias
         as detailed in the text. Solid blue lines show the best-fit power laws, with the $1\sigma$
         uncertainties marked by the shaded regions (see Table \ref{t:scaling}). Different symbols
         identify whether each cluster is disturbed (triangles), relaxed (circles), or not         
         classified (squares). The black dotted line shows the scaling relation found when applying
         the NFW profile correction described in Section \ref{sec:apcorr}. Previous estimates of the
         $Y_{200c}$--$M_{200c}$ scaling relation are shown in the bottom panel with dashed and
         dot-dashed lines (see the text for details).}
\label{fig:scaling} 
\end{figure}

%%%%%%%%%%%%%%%%%%%%%%%%%%%
%%%      Best-fit       %%%
%%%  Scaling Relations  %%%
%%%%%%%%%%%%%%%%%%%%%%%%%%%
\begin{deluxetable}{r c c c c}
\tablecaption{Best-fit Parameters of Scaling Relations}
\tablehead{
\colhead {Relation} &
\colhead{$B_{\rm SS}$\tna} &
\colhead{$A$} &
\colhead{$B$} &
\colhead{$\sigma_{MY}$}
}
\startdata
 $\widetilde{y_0}^{\rm corr}-M_{200c}$ & \ldots & \ytzero & \ytslope & \ytscatter \\[1ex]
       $y_0^{\rm corr}-M_{200c}$       &    1   & \yczero & \ycslope & \ycscatter \\[1ex]
     $Y_{200c}^{\rm corr}-M_{200c}$    &   0.6  &  \Yzero &  \Yslope & \Yscatter
\enddata                                         
\tablecomments{Uncertainties have been estimated through bootstrap resampling.}
\tablenotetext{a}{Expected logarithmic slope from self-similar evolution.}
\label{t:scaling}
\end{deluxetable}

We use the Bivariate Correlated Errors and intrinsic Scatter (BCES) $X_2 \vert X_1$ algorithm for
linear regression \citep{Akritas-96}, which takes into account correlated measurement errors in both
variables and intrinsic scatter, to find the best-fit slopes and normalizations of the power-law
scaling relations given by Equations (\ref{eq:scaling}). The results are shown in Figure
\ref{fig:scaling}, where the solid lines represent the best-fit power laws and the shaded regions
are the $1\sigma$ uncertainties. Table \ref{t:scaling} lists the best-fit parameters, where the last
column lists the log-normal intrinsic scatter orthogonal to the best-fit line, as described by
\cite{Pratt-09}. All uncertainties have been estimated through bootstrap resampling. Different
symbols identify the dynamical state of each cluster (see Section \ref{sec:bias}). It is important
to mention that, in the case of the $Y_{200c}-M_{200c}$ scaling relation, error correlations are
taken into account in the fitting (see also Section \ref{sec:rcorr}).

All three SZE estimators correlate well with dynamical mass, with Pearson's $r$-values of 0.78, 0.82
and 0.86 for $\widetilde{y_0}^{\rm corr}$, $y_0^{\rm corr}$ and $Y_{200c}^{\rm corr}$, respectively.
The fractional errors on the slopes are similar, ranging from 16\% for $\widetilde{y_0}^{\rm
corr}$--$M_{200c}$ to 20\% for $Y_{200c}^{\rm corr}$--$M_{200c}$, while $\sigma_{MY}$ is roughly the
same for all three SZE estimators (but see Section \ref{sec:rcorr}). These values are consistent
with those found in simulations, which have some dependence on the input cluster physics and are of
order 10\%--15\% for large-aperture integrations, such as $Y_{200c}$ \citep{Nagai-06, Yang-10,
Battaglia-12}. We find that the intrinsic scatter of the scaling relations for the central measures
is low and similar to that of $Y_{200c}^{\rm corr}-M_{200c}$, although numerical simulations predict
a higher dependence on gas physics and projection effects for central estimates
\citep[e.g.,][]{Motl-05, Shaw-08}.

As a consistency check, the best-fit power laws have also been estimated using the publicly
available code by \cite{Kelly-07}, which also takes into account measurement errors in both
variables and intrinsic scatter. The scaling relations estimated using this method yield results
that are consistent with those listed in Tables \ref{t:scaling} and \ref{t:mergerbias}, both in
magnitude and uncertainties, for $A$, $B$, and $\sigma_{MY}$.

\subsection{Previous Results}\label{sec:comp_obs}

\cite{Rines-10} were the first to present a comparison of SZE fluxes and masses derived from
dynamical information, but their sample selection did not allow for the estimation of a scaling
relation. Here, we review some SZE-mass scaling relations derived from other observations or mass
proxies. While we note that many authors have presented scaling relations in different forms and
using a variety of mass proxies, here we compare to those that have done so in the same form as is
done here (i.e., correcting by intrinsic evolution in the form of Equations (\ref{eq:scaling})).

When comparing to them, we have converted to values calculated within $r_{200c}$ by assuming that
the mass scales as $M\propto Y^\gamma$, where $\gamma$ is the best-fit slope found in each study.
Specifically, the conversion from a radius $r_\Delta$ to $r_{200c}$ is done by noting that, if
$M_\Delta=\alpha M_{200c}$ (given by an NFW profile) and $Y_\Delta=\beta Y_{200c}$ (given by the A10
profile), then if the scaling relation is of the form $M_\Delta \propto Y_\Delta^\gamma$, it is
straightforward that $\left(\alpha M_{200c}\right) \propto \left(\beta Y_{200c}\right)^\gamma$.

The bottom panel of Figure \ref{fig:scaling} shows these scaling relations; those where masses were
estimated from X-ray observations \citep{Andersson-SPT, Planck-scaling} and weak lensing
measurements \citep{Marrone-12} are shown with dashed lines, and the dash-dotted line shows the
results from hydrodynamical simulations by \cite{Battaglia-12} which include AGN feedback. (The
quoted values for the latter are the results at $z=0.5$, which also corresponds to the
characteristic redshift of our sample.) The latter authors measure $Y_{200c}^{\rm cyl}$, the
integrated Compton parameter within a cylinder of radius $r_{200c}$, which is converted to a
spherical measure following \cite{Arnaud-10}.

The $Y_{200c}^{\rm corr}-M_{200c}$ scaling relation derived in this work is in good general
agreement with the scaling relations cited above, although it is slightly shallower than those
derived by \cite{Andersson-SPT} and \cite{Planck-scaling} and that predicted by
\cite{Battaglia-12}, although the large uncertainties prevent any further analysis. Larger samples
of clusters should help decrease these error bars.

\section{Possible Biases and Systematic Effects}

In this section we explore some effects, both physical and from the analysis, that could be biasing
the results of Section \ref{sec:scaling}. Given the current data set, however, they are all hard to
asses, so we have relied in simulations for some of them. A more detailed treatment of these effects
will be performed in the future, with a larger sample of clusters.

\subsection{Scaling Relations for Relaxed and Disturbed Clusters}
\label{sec:bias}

%%%%%%%%%%%%%%%%%%%%%%%%%%%
%%%     Un/disturbed    %%%
%%%  Scaling Relations  %%%
%%%%%%%%%%%%%%%%%%%%%%%%%%%
\begin{deluxetable}{r c c c c}
\tablecaption{Best-fit Parameters of Scaling Relations for Selected Subsamples}
\tablehead{
\colhead{Relation} &
\colhead{Sample} &
\colhead{$A$} &
\colhead{$B$} &
\colhead{$\sigma_{MY}$}
}
\startdata
\multirow{2}{*}{$\widetilde{y_0}^{\rm corr}-M_{200c}$}  & Disturbed & $14.99\pm0.13$ &
     $0.86\pm0.36$ & $0.15\pm0.09$ \\
                                                             &  Relaxed  & $15.01\pm0.12$ &
     $0.77\pm0.28$ & $0.13\pm0.09$ \\[0.12cm]
\multirow{2}{*}{$y_0^{\rm corr}-M_{200c}$}              & Disturbed & $15.03\pm0.33$ &
     $0.78\pm0.48$ & $0.18\pm0.10$ \\
                                                             &  Relaxed  & $15.01\pm0.11$ &
     $0.93\pm0.20$ & $0.09\pm0.08$ \\[0.12cm]
\multirow{2}{*}{$Y_{200c}^{\rm corr}-M_{200c}$}         & Disturbed & $15.02\pm0.13$ &
     $0.43\pm0.17$ & $0.23\pm0.18$ \\
                                                             &  Relaxed  & $14.96\pm0.11$ &
     $0.58\pm0.19$ & $0.15\pm0.15$
\enddata
\tablecomments{There are seven disturbed and seven relaxed clusters. The scaling relations are in
               the same form, as in Equations (\ref{eq:scaling}). Uncertainties have been estimated
               through bootstrap resampling.}
\label{t:mergerbias}
\end{deluxetable}

Table \ref{t:mergerbias} lists the best-fit scaling relations when separating the sample into
relaxed and disturbed clusters according to Table \ref{t:sub} (see Section \ref{sec:sub}). The
effect of disturbed clusters, if any, is similar for all SZE estimators and is apparent as a slight,
but not significant, change in slope, with disturbed clusters making the slope of the scaling
relations $\sim20\%$ shallower. While errors on samples of this size are very large, we find that
$y_0^{\rm corr}-M_{200c}$ has the largest decrease in scatter when including only relaxed
clusters, and the largest boost for disturbed clusters. However, larger samples of clusters are
needed to test whether disturbed systems induce a significant bias, or larger uncertainties, in the
scaling relations.

As explained in Section \ref{sec:sub}, neither ACT-CL~J0521$-$5104 nor ACT-CL~J0616$-$5227 have been
considered in the present analysis. Including these clusters in either sample does not change the
best-fit parameters and only changes the intrinsic scatter by $\lesssim$0.05, which is within the
quoted uncertainties.

\subsection{Systematic Effects from a Reduced Spectroscopic Coverage}\label{sec:apcorr}

Figure \ref{fig:rv} shows that the spectroscopic coverage does not reach $r_{200c}$ with a
significant number of members in many of the clusters studied here. The spectroscopic aperture is
defined here as the median BCG-centric distance of the last distance-ordered bin of 10 galaxies.
This is assumed to be enough so that a measure of the velocity dispersion of these galaxies at such
distance is representative of all member galaxies (with and without a redshift measurement) in this
bin. The distribution of apertures is asymetric, with an average $r_{\rm ap}=0.55_{-0.24}^{+0.36}$
(90th and 10th percentiles), and is a function of the angular diameter distance $D_A(z)$ and the
size of the cluster (hence the mass). Thus, more massive clusters at lower redshift have the
smallest coverage.

Observations and simulations seem to give different answers to what should the velocity dispersion
profile of a cluster look like. Simulations show that the velocity dispersion profile for an NFW
density profile should be decreasing with radius \citep[e.g.,][]{Biviano-06,Mamon-10}. Most
observations, however, find that, on average, the velocity dispersion profile of clusters is flat
outside $r\approx0.5r_{200c}$ \citep[e.g.,][]{Biviano-03,Katgert-04,Faltenbacher-06,Lokas-06},
although some observations do support the expectations from simulations \citep[e.g.,][]{Rines-03}.

Because there are many unknowns in the size of the correction and the cluster properties that drive
it, we do not correct our mass measurements by any bias introduced by this reduced coverage.
However, we do estimate what the bias could be based on theoretical predictions. We use the velocity
dispersion profile predicted by an NFW profile, as derived by \cite{Mamon-10} from {\it N}-body
simulations, using the mass--concentration relation of \cite{Duffy-08}. We correct our measurement
to
a measurement at $r_{200c}$ using this profile and measure a ``corrected'' $M_{200c}$. Since the
re-calculation of $r_{200c}$ for a lower (higher) mass means that we have under(over-)estimated the
actual sampling aperture, this procedure is iterated until the radius converges, which takes 3--4
iterations. The average correction to the velocity dispersion is 0.91, and the average mass
correction derived from Equation (\ref{eq:mass}) is 0.79. The correction to $r_{200c}$ is of the
same order as that of the velocity dispersion, and we use these corrected radii to estimate
corrected $Y_{200c}$, which are on average 0.91 of those reported in Table \ref{t:sze}. As mentioned
in Section \ref{sec:sze_par}, $\widetilde{y_0}$ and $y_0$ are measured completely independent from
the dynamical masses, so these values are not affected. The scaling relations estimated from the
corrected numbers are shown in each panel of Figure \ref{fig:scaling} as a black dotted line. The
effect of the correction is to flatten the slopes and lower the normalizations, with
$A=\{14.93\pm0.06,14.83\pm0.06,14.91\pm0.06\}$ and $B=\{0.60\pm0.09,0.66\pm0.08,0.40\pm0.09\}$ for
$\{\widetilde{y_0},y_0,Y_{200c}\}$. The resulting scatter is slightly lower but consistent with the
values reported in Table \ref{t:scaling} for all estimators.

We also estimated, for comparison, the correction obtained when applying the surface pressure
correction term \citep{The-86,Girardi-98}, assuming an NFW profile. This correction is directly
applied to the mass measurement. The average correction to the present sample is 0.73. This yields
best-fit scaling relations with shallower slopes but consistent with the previous method. However,
this correction is applicable specifically to the virial mass measurement (i.e., estimated from the
Virial Theorem), so should be taken with care, especially for a population that may be dominated by
dynamically disturbed clusters like the present one.

\subsection{The Redshift Evolution of $\widetilde{y_0}$}

While the functional forms of Equations (\ref{eq:scaling}) are well motivated from self-similar
evolution for $y_0$ and $Y_{200c}$, $\widetilde{y_0}$ is dependent on the adopted filtering of the
maps \citep{Hasselfield-ACT} and we have no a priori information on how it should evolve with 
redshift for a fixed mass. We have explored a range of functional forms for the redshift dependence
of Equation (\ref{eq:ytilde}) using the set of models discussed in Section \ref{sec:corrections}. We
find that, while the results are consistent, the models prefer a slightly lower redshift evolution
of $\widetilde{y_0}$ at fixed mass. Specifically, the scaling resulting from Equation
(\ref{eq:ytilde}) could, at low ($z\sim0.3$) redshifts, bias the masses high (on average) by as much
as $\sim25\%$. Conversely, at high ($z\sim1$) redshifts the masses could be biased low by up to
$\sim35\%$.

These new mass predictions would, for a variety of redshift parameterizations and for all the
clusters in our sample, be within the measured $1\sigma$ uncertainties. Given the sample size and
measurement uncertainties, we have decided to study $\widetilde{y_0}$ in a similar way to $y_0$, to
be able to compare more easily the two, which are closely related. As mentioned above, a more
detailed study of the functional form of Equations (\ref{eq:scaling}) will be performed in future
work with a larger sample of clusters.

\subsection{$M_{200c}$--$Y_{200c}$ Correlation}\label{sec:rcorr}

Since $r_{200c}$ is derived from dynamical information and used to estimate $Y_{200c}$, there is
non-zero correlation between $Y_{200c}$ and $M_{200c}$. As noted in Section \ref{sec:fits}, the
best-fit slopes and normalizations listed in Tables \ref{t:scaling} and \ref{t:mergerbias} include
error correlations between these two parameters. The effect of this correlation is to flatten the
relation notably, although within errorbars: if not included in the BCES fit, the slope increases to
$0.56\pm0.11$.

Additionally, as discussed analogously by \cite{Becker-11} in the context of the $M_{500c}-M_{\rm
gas}$ relation, such a correlation will bias the intrinsic scatter measurement low by a factor
$\approx1-\alpha/3$, where $Y_{200c} \propto (r/r_{200c})^\alpha$ near $r_{200c}$. By re-calculating
$Y_{200c}$ at different radii around $r_{200c}$ (see Section \ref{sec:sze_par}), we find
$\alpha\simeq1.18$. The intrinsic scatters of the $Y_{200c}-M_{200c}$ relations in Tables
\ref{t:scaling} and \ref{t:mergerbias} consequently include a correction factor $\approx1.65$, which
makes them larger than the intrinsic scatters for the other relations but consistent within the
large errorbars.

\section{Discussion}

\subsection{Individual Clusters}\label{sec:previous}

In this section we list clusters with notable features, including comparison of dynamical masses
presented here with previous estimates, where available.\footnote{We quote the original mass
estimates, given as $M_{500c}$---also with respect to the {\it critical} density of the
universe---and assume a typical conversion factor $M_{200c} \approx 1.6\,M_{500c}$
\citep{Duffy-08} when comparing with our results. In the particular case of ``El Gordo'', masses
are originally given as $M_{200a}$, the masses within a radius containing 200 times the {\it
average} density of the universe. For this cluster, the conversion is
$M_{200c}\approx0.86M_{200a}$.} The respective original or alternative names can be found in Table
\ref{t:dyn}. With respect to notes on optical features of these clusters, the reader is referred to
Figures 4--10 of \cite{Menanteau-ACT}, as appropriate.

\subsubsection{ACT-CL~J0102$-$4915 ``El Gordo''}
Located at $z = 0.870$, this cluster has the largest SZE signal of all ACT clusters (it is the
rightmost data point in all panels of Figure \ref{fig:scaling}) and is one of the most massive
clusters of the sample according to its dynamics. This cluster looks elongated in the optical (in
fact, it is double-peaked in the galaxy distribution; \citealt{Menanteau-J0102}), but there are no
clear signs of l.o.s. substructure from the dynamical information. In \cite{Menanteau-J0102},
we show that if the cluster is divided into two subclusters in the process of merging (as suggested
by the optical data), they have a mass ratio of order 2:1, with a total summed dynamical mass of
$M_{200c} = (24 \pm 7) \times10^{14}\,h_{70}^{-1}\,M_\odot$, making this a huge merger between two
already massive clusters.

\cite{Menanteau-J0102} used a multi-wavelength data-set combining X-rays, SZE and the information
provided in this work to estimate the cluster mass using several mass proxies which are in
statistical agreement, with a combined mass estimate of $M_{200a} =
21.6\pm3.2\times10^{14}\,h_{70}^{-1}\,M_\odot$. The statistical error for the combined mass is
likely an underestimate given the complex nature of this massive merging cluster \citep[see][for a
detailed discussion of the mass measurements in ``El Gordo'']{Menanteau-J0102}

\subsubsection{ACT-CL J0215$-$5212}
As shown in Table \ref{t:sub}, this cluster appears to have substructure with a high significance as
given by the DS test. More noteworthy, however, is the peculiar velocity of the BCG, $v_{\rm pec} =
1171\pm153\,\mathrm{km\,s^{-1}}$, different from zero at $>7.5\sigma$. This is the only cluster in
our sample in which the velocity of the BCG is comparable to the velocity dispersion of the cluster,
and the cluster where the emission-line galaxies are most different from the whole population.
ACT$-$CL~J0215-5212 has a second galaxy $\sim23\arcsec$ away (corresponding to a projected distance
of $140\,h_{70}^{-1}\,\rm{kpc}$ at $z=0.480$) which is only 0.27 mag fainter and has a peculiar
velocity of roughly $-660\,\mathrm{km\,s^{-1}}$, and at least three more galaxies within 0.55 mag of
the BCG (which is the brightest of all by definition, but is also the one nearest to the optical
center of the cluster), all of which have comparable ($\sim1000\,\mathrm{km\,s^{-1}}$) peculiar
velocities. On the other hand, this cluster does not significantly depart from any of the scaling
relations of Figure \ref{fig:scaling}, showing the complexity of substructure analyses. It also has
the highest fraction of emission-line galaxies.

\subsubsection{ACT-CL~J0237$-$4939}\label{sec:previous-J0237}
Similar to the previous case, this cluster has three bright galaxies within 2$\arcmin$ of the BCG,
which are within 1 mag of the BCG. In particular, the second-brightest galaxy is 65$\arcsec$
($310\,h_{70}^{-1}\,$ kpc) away from the BCG, is 0.49 mag fainter and has a peculiar velocity with
respect to the cluster of $\sim1850\,\mathrm{km\,s^{-1}}$. All this argues in favor of the
classification of this cluster as a disturbed system.

\subsubsection{ACT-CL~J0330$-$5227}
As mentioned in Section \ref{sec:sample}, this cluster was discovered by \cite{Werner-07} behind
A3128 ($z = 0.06$) using \xmm\ X-ray observations. The SZE measurement is clearly associated with
the background structure while the less massive, foreground cluster has no significant SZE emission
\citep{Hincks-ACT}. The dynamical mass estimated here is significantly higher than that derived by
\cite{Werner-07}, of $M_{500c} = 3.4\times10^{14}\,h_{70}^{-1}\,M_\odot$. They do caution, however,
that their estimate is uncertain, as it is based on an isothermal beta-model for the cluster. Being
located only 12\arcmin\ away from A3128 at $z=0.06$ on the sky, this cluster is a clear
illustration of the mass selection of SZE surveys, approximately independent of redshift
\citep[see][for further discussion]{Hincks-ACT}.

\subsubsection{ACT-CL~J0438$-$5419}
ACT-CL~J0438$-$5419 is the only new ACT cluster also reported by the \planck\ satellite in its early
release \citep[PLCKESZ G262.7$-$40.9;][]{Planck-clusters}.\footnote{Five other clusters in this
sample (El Gordo, ACT-CL~J0235$-$5121, ACT-CL~J0304$-$4921, ACT-CL~J0559$-$5249 and
ACT-CL~J0707$-$5522) have now been included in the Planck SZ catalog \citep{Planck-PSZ}.} It has
been followed-up with \xmm, with which \cite{Planck-xmm} estimated a mass $M_{500c} = (6.9 \pm
0.7)\times10^{14} \,h_{70}^{-1}\,M_\odot$ using a $Y_X$--$M$ scaling. This value is $1.6\sigma$
lower than our dynamical estimate; in fact, this cluster is one of the most massive ones in our
sample. However, their reported errors include only statistical effects, so are underestimates of
the true errors in the measurement.

This cluster is also reported in \cite{Williamson-11}. They estimate a simulation-based
SZE-estimated mass $M_{500c} = (8.2 \pm 2.5)\times10^{14} \,h_{70}^{-1}\,M_\odot$, consistent with
our dynamical mass estimate.

\subsubsection{ACT-CL~J0509$-$5341}
This was one of the first clusters discovered by SPT \citep{Staniszewski-SPT} and the first mass
measurements were reported by \cite{Menanteau-09}. \cite{Vanderlinde-SPT} reported a
simulation-based SZE estimate of the mass of $M_{500c} = (4.3 \pm
1.1)\times10^{14}\,h_{70}^{-1}\,M_\odot$ and \cite{Andersson-SPT} estimated an X-ray $Y_X$--derived
mass from $M_{500c} = (5.4 \pm 0.6)\times10^{14}\,h_{70}^{-1}\,M_\odot$. All previous values are in
agreement with our estimate. Consistent with our substructure analysis, \cite{Andersson-SPT} found
that this cluster is a disturbed system based solely on X-ray morphology.

\subsubsection{ACT-CL~J0521$-$5104}
This cluster is not reported in \cite{Marriage-ACT}, because it was not an S/N$>$3 detection in the
initial analysis. However, more recent analyses including data from 3 yr of observations show
that this cluster is now detected at $4.5\sigma$, and it is therefore included in this study.
\cite{Vanderlinde-SPT} report an SZE-estimated mass of $M_{500c} =
(2.97\pm0.89)\times10^{14}h_{70}^{-1}\,M_\odot$, significantly lower than the dynamical mass
reported here.

\subsubsection{ACT-CL~J0528$-$5259}
This cluster was also reported by \cite{Staniszewski-SPT} and characterized optically by
\cite{Menanteau-09} Its SZE-estimated mass is $M_{500c} = (2.9 \pm
0.9)\times10^{14}\,h_{70}^{-1}\,M_\odot$ \citep{Williamson-11} and its X-ray-estimated mass is
$M_{500c} = (3.0 \pm 0.9)\times10^{14}\,h_{70}^{-1}\,M_\odot$ \citep{Andersson-SPT}. These values
are consistent with our dynamical estimate. Also consistent with the present finding,
\cite{Andersson-SPT} found that the X-ray morphology shows this cluster to be relaxed.

\subsubsection{ACT-CL~J0546$-$5345}\label{sec:previous-J0546}
This is the highest-redshift cluster of the sample, at $z = 1.066$. \cite{Brodwin-SPT} first
presented a spectroscopic study of this cluster based on 18 cluster members, which have been
included in this study, plus the three emission-line galaxies not used for their mass measurement.
We included \citeauthor{Brodwin-SPT}'s (\citeyear{Brodwin-SPT}) galaxies in our spectroscopic
catalog and applied the cluster membership algorithm (Section \ref{sec:vdisp}) which resulted in 48
members in total. Our mass estimate, calculated now with three times as many galaxies, is consistent
both with their dynamical estimate and their best estimate, combining X-ray, SZE and dynamical
information, which corresponds to $M_{200c} = (7.9 \pm 0.9) \times10^{14}\,h_{70}^{-1}\,M_\odot$.

\subsubsection{ACT-CL~J0559$-$5249}
This cluster was also detected by SPT and reported in \cite{Vanderlinde-SPT}. They report a
simulation-based SZE-derived mass $M_{500c} = (5.3 \pm 1.2)\times10^{14} \,h_{70}^{-1}\,M_\odot$,
while \cite{Andersson-SPT} estimate an X-ray $Y_X$-derived mass of $M_{500c} = (6.4 \pm
0.5)\times10^{14} \,h_{70}^{-1}\,M_\odot$. Both these estimates are consistent with each other, and
combined are consistent with the lower limit of our dynamical estimate. The ACT SZE signal is
consistent with the dynamical mass (cf.\ Figure \ref{fig:scaling}). Using X-ray data,
\cite{Andersson-SPT} suggest that this cluster is in the process of merging but our substructure
analysis finds no evidence for substructure. These two results are not necessarily in contradiction
since X-ray morphology and dynamical information are sensitive to substructure with different
orientations.

\subsubsection{ACT-CL~J0616$-$5227}
The optical imaging of this cluster by \cite{Menanteau-ACT} was sufficient to provide confirmation
but shallower than required to secure an adequate galaxy catalog for spectroscopic targeting. Out of
73 spectra obtained, only 18 are cluster members. Another six are foreground/background galaxies.
The remaining are all late-type (mostly M) stars, which have similar colors to the cluster members.
Both the SZE signal and the X-rays argue in favor of this being a massive cluster, supporting the
dynamical estimate.

\subsection{The SCS Clusters}\label{sec:SCS}

%%%%%%%%%%%%%%%%%%%%%%%%%%%
%%  SCS clusters Table  %%%
%%%%%%%%%%%%%%%%%%%%%%%%%%%
\begin{deluxetable*}{c c c c c c}
\tablecaption{Clusters in the Optical Program not Netected by ACT in the 2008 Observing Season.}
\tablehead{
\colhead{Name} &
\colhead{$N_{\rm gal}$} &
\colhead{$z_{\rm BI}$} &
\colhead{$S_{\rm BI}$} &
\colhead{$r_{200c}$} &
\colhead{$M_{200c}$}
 \\
\colhead{} &
\colhead{} &
\colhead{} &
\colhead{$(\mathrm{km\,s^{-1}})$} &
\colhead{$(h_{70}^{-1}\,\mathrm{kpc})$} &
\colhead{$(10^{14}\,h_{70}^{-1}\,M_\odot)$}
}
\startdata
SCSO J0514$-$5126 & 15 & $0.7372\pm0.0018$ & \phn$931\pm154$    &    $1370\pm218$ & $6.3\pm3.0$\\
SCSO J0514$-$5140 & 22 & $0.7362\pm0.0011$ & \phn$701\pm125$    &    $1036\pm182$ & $2.7\pm1.4$\\
SCSO J0540$-$5614 & 17 & $0.4477\pm0.0008$ & \phn$578\pm115$    & \phn$990\pm193$ & $1.7\pm1.0$
\enddata
\label{t:SCS}
\end{deluxetable*}

Of the 19 clusters observed during this program (see Table \ref{t:runs}), only the 16 listed in
Table \ref{t:dyn} were detected by ACT. The other three clusters are listed in Table \ref{t:SCS}.
These three clusters (hereafter ``the SCS clusters'') were discovered optically in the SCS and were
included in the spectroscopic program because of their high optical richness \citep{Menanteau-10},
along with ACT-CL~J0521$-$5104. Despite them being optically rich systems, the masses of the three
SCS clusters are consistent with being below the ACT detection limit.

\section{Conclusions}

We have conducted a large spectroscopic follow-up program of clusters of galaxies discovered via the
SZE by ACT in its southern sky survey \citep{Menanteau-ACT, Marriage-ACT}. We
used 89 hr of multi-object spectroscopic observations divided between FORS2 at VLT and GMOS at
Gemini-South. With a few (3--4) hr of observation per cluster, we have been able to confirm an
average 65 members per cluster, which allowed us to: (1) obtain robust redshifts for each cluster;
(2) measure velocity dispersions with errors $\lesssim10\%$, which translates to uncertainties of
$<30\%$ in mass estimates; and (3) determine the dynamical state of the clusters.

The cluster sample spans a redshift range $0.28 < z < 1.07$, with a median redshift $z = 0.50$.
Careful examination of possible substructure shows that $\sim50\%$ of the clusters in the ACT sample
show signs of significant substructure, consistent with the X-ray study of SPT SZE-selected clusters
\citep{Andersson-SPT} and with optically and X-ray-selected local clusters. We find that the
presence of emission-line galaxies within clusters, which could be associated with infalling groups,
does not significantly modify the mass estimates. For this reason, emission-line galaxies have been
included as members in the final samples.

Dynamical masses have been estimated from the radial velocity dispersions using the \cite{Evrard-08}
simulation-based $\sigma-M_{200c}$ scaling relation. These clusters have masses $6 \lesssim M_{200c}
\lesssim 21$ in units of $10^{14}\,h_{70}^{-1}\,M_\odot$, with a median mass
$\sim12\times10^{14}\,h_{70}^{-1}\,M_\odot$ in agreement with the mass distribution of the ACT
sample as estimated from X-ray luminosities \citep{Marriage-ACT}. These clusters rank therefore
among the most massive clusters in the universe.

The scaling relation between dynamical mass and SZE signal has been studied using three estimators
of the SZE: the central match-filtered SZE amplitude, $\widetilde{y_0}$, the central Compton
parameter, $y_0$, and the Compton signal integrated within $r_{200c}$, $Y_{200c}$. In order to
derive unbiased scaling relations, a simulation-based flux bias correction has been applied to the
data, and the scaling relations include intrinsic evolution with redshift.

These scaling relations are summarized in Table \ref{t:scaling} and represent the main result of
this work. The intrinsic scatter in these relations is consistent with that predicted by simulations
\citep[e.g.,][]{Motl-05, Reid-06}. We find that all our SZE estimators are similarly robust as mass
proxies, with lognormal intrinsic scatters $\sim15\%$, although in the case of $Y_{200c}-M_{200c}$,
the scatter is high but poorly constrained because of correlations between the observables. The
derived scaling relations agree with the expectations from self-similar evolution of clusters.
Although there are hints that dynamically disturbed clusters may bias the scaling relations, the
present sample size does not allow for a robust constrain on this effect. The scaling relation
between $Y_{200c}$ and dynamical mass $M_{200c}$ is in good agreement with previous results, which
are based on different mass proxies, and predictions from simulations.

In summary, the first sample of spectroscopically followed-up SZE-selected clusters from ACT has
yielded results that agree with the expectations for the first-generation of SZE surveys. The
scaling relations derived from this sample also agree with the expectations. The results presented
here show that dynamical masses provide a good way of relating the SZE to cluster masses, and larger
cluster samples, in combination with other mass proxies, will serve as a tight constraint for
cosmology.

\acknowledgments

% Personal
We are grateful to Andrea Biviano for lengthy discussions on the corrections to the velocity
dispersion, to Gary Mamon for kindly providing the velocity dispersion profiles obtained from their
simulations and to Gus Evrard for helpful discussions on the aperture correction, and to Gabriel
Pratt and Gilbert Holder for useful comments on the original draft. We also thank the anonymous
referee for a very insightful revision of the successive versions of the draft, which helped improve
the consistency and robustness of this work.

% ACT
ACT operates in the Parque Astron\'omico Atacama in northern Chile under the auspices of Programa de
Astronom\'ia de la Comisi\'on Nacional de Investigaci\'on Cient\'ifica y Tecnol\'ogica de Chile
(CONICYT).
% NSF
This work was supported by the U.S. National Science Foundation through awards AST-0408698 and
AST-0965625 for the ACT project, and PHY-0855887, PHY-1214379, AST-0707731 and PIRE-0507768 (award
No.\ OISE-0530095).
% Princeton/Rutgers
Funding was also provided by Princeton University, the University of Pennsylvania, and a Canada
Foundation for Innovation (CFI) award to UBC.
% SciNet
Computations were performed on the GPC supercomputer at the SciNet HPC Consortium. SciNet is funded
by the CFI under the auspices of Compute Canada, the Government of Ontario, the Ontario Research
Fund -- Research Excellence; and the University of Toronto.
% PUC
This research is partially funded by ``Centro de Astrof\'isica FONDAP'' 15010003, Centro BASAL-CATA,
by FONDECYT under projects 1120676 and 1085286 and by ALMA-CONICYT under projects 31090002 and
31100003.

\pagebreak

\begin{appendix}

\section{Electronic Data}

Table \ref{t:catalog} lists the properties of the BCGs for each of the ACT clusters (see Table
\ref{t:dyn}). This table is an excerpt from the full table available online (from which the BCGs are
shown for convenience), which contains all cluster members for the 16 ACT clusters. It is given for
guidance in its form and content. Column 1 lists the adopted identification, based on the J2000.0
position of each galaxy and using the initials of the first three authors of this paper to identify
the catalog. Columns 2 and 3 list the positions of the galaxies. Column 4 lists the magnitude in
the $i$ band and Column 5 lists the cross-correlation redshifts and their associated errors as given
by RVSAO. Column 6 lists the cross-correlation S/N $r_{cc}$ \citep{Tonry-79} and Column 7 lists the
main spectral features of each galaxy.

\begin{deluxetable*}{c c c c c r l}[b]
\tablecaption{Spectroscopic Members of the 16 ACT Clusters}
\tablehead{
\colhead{Identification} &
\colhead{R.A.} &
\colhead{Decl.} &
\colhead{$m_i$} &
\colhead{$z$} &
\colhead{$r_{cc}$} &
\colhead{Main Spectral}
 \\
\colhead{} &
\colhead{(hh:mm:ss)} &
\colhead{(dd:mm:ss)} &
\colhead{} &
\colhead{} &
\colhead{} &
\colhead{Features}
}
\startdata
SMH\_J010257.7$-$491619.2 & 01:02:57.74 & $-$49:16:19.2 & 19.186 & $0.87014\pm0.00030$ &  3.39 &
Ca {\sc ii} K,H; [O {\sc ii}]\\
SMH\_J021512.3$-$521225.3 & 02:15:12.26 & $-$52:12:25.3 & 18.678 & $0.48587\pm0.00016$ &  3.90 &
Ca {\sc ii} K,H \\
SMH\_J023242.8$-$525722.3 & 02:32:42.80 & $-$52:57:22.3 & 18.410 & $0.55592\pm0.00014$ &  4.53 &
Ca {\sc ii} K,H \\
SMH\_J023545.3$-$512105.2 & 02:35:45.28 & $-$51:21:05.2 & 16.493 & $0.27825\pm0.00015$ &  7.18 &
Ca {\sc ii} K,H \\
SMH\_J023701.7$-$493810.0 & 02:37:01.71 & $-$49:38:10.0 & 17.582 & $0.33554\pm0.00016$ & 10.42 &
Ca {\sc ii} K,H \\
SMH\_J030416.0$-$492126.3 & 03:04:16.04 & $-$49:21:26.3 & 17.463 & $0.39289\pm0.00020$ &  9.43 &
Ca {\sc ii} K,H \\
SMH\_J033056.8$-$522813.7 & 03:30:56.83 & $-$52:28:13.6 & 17.520 & $0.43969\pm0.00019$ & 10.23 &
Ca {\sc ii} K,H \\
SMH\_J034655.5$-$543854.8 & 03:46:55.49 & $-$54:38:54.8 & 18.577 & $0.53107\pm0.00013$ &  6.16 &
Ca {\sc ii} K,H \\
SMH\_J043817.7$-$541920.7 & 04:38:17.70 & $-$54:19:20.6 & 17.470 & $0.41955\pm0.00012$ &  9.42 &
Ca {\sc ii} K,H \\
SMH\_J050921.3$-$534212.2 & 05:09:21.38 & $-$53:42:12.2 & 18.361 & $0.46257\pm0.00022$ &  7.53 &
Ca {\sc ii} K,H; [O {\sc ii}]\\
SMH\_J052114.5$-$510418.5 & 05:21:14.54 & $-$51:04:18.6 & 19.060 & $0.67780\pm0.00041$ &  3.96 &
Ca {\sc ii} K,H \\
SMH\_J052805.3$-$525952.8 & 05:28:05.30 & $-$52:59:52.8 & 19.715 & $0.76695\pm0.00037$ &  6.10 &
Ca {\sc ii} K,H \\
SMH\_J054637.6$-$534531.3 & 05:46:37.67 & $-$53:45:31.3 & 21.184 & $1.06255\pm0.00016$ &  6.47 &
Ca {\sc ii} K,H \\
SMH\_J055943.2$-$524927.1 & 05:59:43.23 & $-$52:49:27.1 & 19.103 & $0.61035\pm0.00027$ &  3.88 &
Ca {\sc ii} K,H \\
SMH\_J061634.1$-$522709.9 & 06:16:34.05 & $-$52:27:09.9 & 18.594 & $0.68765\pm0.00011$ &  6.87 &
Ca {\sc ii} K,H \\
SMH\_J070704.7$-$552308.4 & 07:07:04.67 & $-$55:23:08.4 & 16.754 & $0.29451\pm0.00019$ &  6.05 &
Ca {\sc ii} K,H
\enddata
\tablecomments{BCGs of the 16 SZE-selected clusters presented here. Galaxies have been named based
on their positions, and using the initials of the first three authors of this paper to identify the
catalog. \\ (This table is available in its entirety in a machine-readable form in the online
journal. A portion is shown here for guidance regarding its form and content.)}
\label{t:catalog}
\end{deluxetable*}

\end{appendix}

\end{document}